\documentclass[useAMS,usenatbib]{mn2e}
\usepackage{graphicx}
\usepackage{epsfig}
\usepackage{verbatim}
\usepackage[flushleft]{threeparttable}
\usepackage{amssymb}
\usepackage{amsmath}

\def\deg{{$^\circ$}} 
\def\apjl{ApJL}
\def\apj{ApJ}
\def\apjs{ApJS}
\def\mnras{MNRAS}
\def\araa{ARAA}
\def\aj{AJ}
\def\aap{A\&A}
\def\aaps{A\&A Suppl.}
\def\nat{Nature}

\newcommand{\etal}{et~al.\ }

\title[Galaxy Properties Across Projected Phase Space in A901/2]{OMEGA -- OSIRIS Mapping of Emission-line Galaxies in
A901/2: III. -- Galaxy Properties Across Projected Phase Space in A901/2}

\author[Weinzirl~et.~al.]{Tim Weinzirl$^{1}$\thanks{E-mail:
timothy.weinzirl@nottingham.ac.uk}, Alfonso~Arag\'on-Salamanca$^{1}$\thanks{E-mail:
alfonso.aragon@nottingham.ac.uk}, Meghan~E.~Gray$^{1}$, 
\newauthor Steven~P.~Bamford$^{1}$, Bruno~Rodr\'iguez~del~Pino$^{1,2}$,  Ana~L.~Chies-Santos$^{3}$,
\newauthor Asmus B\"ohm$^{4}$, Christian Wolf$^{5}$, Richard J. Cool$^{6,7}$\\
$^{1}$School of Physics and Astronomy, The University of Nottingham, University Park, Nottingham, NG7 2RD, UK\\
$^{2}$Centro de Astrobiolog\'ia, INTA-CSIC, Madrid, Spain\\
$^{3}$Departamento de Astronomia, Instituto de Física, Universidade Federal do Rio Grande do Sul, Porto Alegre, R.S, Brazil\\
$^{4}$Institute for Astro- and Particle Physics, University of Innsbruck, Technikerstr. 25/8, A-6020 Innsbruck, Austria\\
$^{5}$Research School of Astronomy and Astrophysics, Australian National University, Cotter Road, Weston Creek, ACT 2611, Australia\\
$^{6}$ MMT Observatory, Tucson, AZ 85721, USA\\
$^{7}$ Netflix, 100 Winchester Circle, Los Gatos CA, 95032\\
}
\begin{document}

\date{Accepted 2017 June 15. Received 2017 June 15; in original form 2017 March 3}

\pagerange{\pageref{firstpage}--\pageref{lastpage}} \pubyear{2017}

\maketitle

\label{firstpage}

\begin{abstract}
We conduct a comprehensive projected phase-space analysis of the A901/2 multi-cluster system at 
$z\sim0.165$. Aggregating redshifts from spectroscopy, tunable-filter imaging, and prism 
techniques, we assemble a sample of 856 cluster galaxies reaching 
$10^{8.5}\,M_\odot$ in stellar mass. We look for variations in cluster galaxy 
properties between virialised and non-virialised regions of projected phase-space (PPS).
Our main conclusions point to 
relatively gentle environmental effects, expressed mainly on galaxy gas reservoirs. 
(1)~Stacking the four subclusters in A901/2, we find galaxies in the virialised 
region are more massive, redder, and have marginally higher S\`ersic indices, but
their half-light radii and Hubble types are not significantly different. 
(2)~After accounting for trends in stellar mass, there is a remaining change in rest-frame 
colour across PPS. Primarily, the colour difference is due to an 
absence in the virialised region of galaxies with rest-frame $B-V<0.7$ and 
moderate-to-high ($M_\star>10^{9.85}$~$M_\odot$) stellar mass.
(3)~There is an infalling population of lower-mass 
($M_\star\leq10^{9.85}$~$M_\odot$), relatively blue ($B-V<0.7$) 
elliptical or spheroidal galaxies that is strikingly absent in the virialised region.
(4)~The number of bona-fide star-forming and AGN galaxies in the PPS regions is
strongly dictated by stellar mass.
However, there remains a reduced fraction of star-forming galaxies in the centres of 
the clusters at fixed stellar mass, consistent with the star-formation-density relation 
in galaxy clusters.
(5)~There is no change in specific H$\alpha$-derived star-formation rates of star-forming galaxies
at fixed mass across 
the cluster environment. This suggests that preprocessing of galaxies during infall 
plays a prominent role in quenching star formation.
\end{abstract}
\begin{keywords}
galaxies: distances and redshifts -- galaxies: clusters -- astronomical instrumentation: interferometers
\end{keywords}

\section{Introduction}
In hierarchical models of galaxy evolution, observable properties of galaxies, such as
structure and star formation history, are determined by different mass assembly
mechanisms, secular processes, and environment. In particular, hierarchical growth
proceeds through gas and stars assembled via major (e.g., Toomre 1977; Barnes 1988;
Khochfar \& Silk 2006, 2009; Hopkins et al. 2009) and minor (e.g., Hopkins
et al. 2009; Oser et al. 2012; Hilz et al. 2013) mergers with other galaxies,
as well as gas accretion from the halo and cold streams (e.g., Birnboim \& Dekel 2003;
Dekel \& Birnboim 2006; Kere{\v s} et al. 2005; Ocvirk et al. 2008).
Bar-driven gas inflows can efficiently redistribute angular
momentum and mass, and drive internal secular evolution
 (e.g., Kormendy 1993; Kormendy \& Kennicutt 2004; Jogee et al. 2005).
In high-density environments (e.g., rich galaxy clusters),
physical processes such as
ram-pressure stripping (Gunn \& Gott 1972; Fujita \& Nagashima 1999), galaxy
harassment (Barnes \& Hernquist 1991; Moore et al. 1996; Hashimoto
et al. 1998; Gnedin 2003), and strangulation (Larson et al. 1980) can alter star formation
history and galaxy stellar structure.

Yet, after over a century of exploring galaxies, we are still grappling with questions of how 
galaxies form and evolve in different environments across cosmic time.
It is well known that strong observational differences exist between the morphology,
colour, and star formation of galaxies in field-like environments and in rich clusters,
particularly at low ($z\lesssim1$) redshifts  (e.g., Dressler 1980;
Butcher \& Oemler 1984;  Poggianti \etal 2001, 2008;
Postman et al. 2005;  Tran et al. 2005; Cooper 2007, 2008; Lidman \etal 2008;   Patel \etal 2009;
Bauer \etal 2011). For instance, present-day clusters are dominated by galaxies
with elliptical or S0 morphologies, while spirals dominate low-density, field-like
environments (e.g., Dressler 1980; Postman \& Geller 1984; Norberg et al. 2002; Goto et al. 2003; Blanton et al. 2005; 
Postman et al. 2005; Desai et al. 2007; Wolf et al. 2007; Ball, Loveday \& Brunner 2008).

Most previous studies of galaxy environment are two-dimensional and subject to large uncertainties from projection 
effects, particularly in clusters where environmental effects are expected to be strongest. 
Truly understanding the role of galaxy environment on galaxy evolution requires a three-dimensional analysis that
considers location in the ``projected'' phase-space (PPS) of cluster-centric radius and line-of-sight velocity. 

Recent simulations 
(e.g., Mahajan, Mamon \& Raychaudhury 2011; Oman, Hudson \& Behroozi 2013) 
have demonstrated that a galaxy's 
orbital history is predictive of the region in phase-space that it occupies. In particular, infalling
galaxies can be distinguished from galaxies in the virialised cluster core.  Thus, it is possible to infer the 
assembly history of galaxy clusters by correlating phase-space to other observable galaxy properties.

In the last few years, galaxy phase-space has been exploited by several authors in efforts to understand 
galaxy properties like star formation and gas stripping (Hern{\'a}ndez-Fern{\'a}ndez et al. 2014), quenching 
timescales (Muzzin et al. 2014; Taranu et al. 2014), HI stripping in galaxies
(Jaff\'e et al. 2015), and galaxy dust temperatures (Noble et al. 2016). 

Here, we present our own study of phase-space in the A901/2 multi-cluster system at $z\sim0.165$. 
This study jointly exploits two complementary observing campaigns focused on A901/2, the OSIRIS 
Mapping of Emission-line Galaxies in A901/2 (OMEGA) survey (Chies-Santos et al. 2015; 
Rodr\'iguez del Pino et al. 2017), which measured redshifts from the 
H$\alpha$ and {\rm [N\textsc{ii}]} emission lines of galaxies in A901/2, and
the Space Telescope A901/2 Galaxy Evolution Survey (STAGES, Gray et al. 2009). 
These surveys yield complementary information about galaxy
structure and star formation history. 

In Section~\ref{data}, we combine multiple galaxy redshift surveys to assemble a 
representative sample of 856 galaxies complete in stellar mass down to 
$10^{8.5}$~$M_\odot$. In Section~\ref{sphasespace} we outline our phase-space 
analysis. In Section~\ref{dependence}, we investigate how galaxy properties 
vary across phase-space and stellar mass. By comparing galaxy properties in regions of 
the PPS that statistically separate the infalling population from the population of the 
virialised region, we can attempt to disentangle which physical processes are 
influencing galaxies at various points during their accretion history onto the clusters.
We discuss the broader implications of our results in Section~\ref{sdiscuss}, and in 
Section~\ref{ssummary} we summarise our findings.

We adopt a flat $\Lambda$ cold dark matter cosmology with $\Omega_\Lambda=0.7$ and 
$H_0$~=~70~km~s$^{-1}$ Mpc$^{-1}$. 
We use Vega magnitudes throughout the paper.

\section[]{Data and Sample Selection}\label{data}
{The multi-wavelength STAGES survey covers the A901/2 system of four
subclusters at $z\sim 0.165$.  The available photometry includes the 17-band Classifying Objects by Medium-Band Observations 
survey (COMBO-17; Wolf et al. 2003), covering $31.5 \times 30.0\,$arcmin$^2$ ($\sim5.3\times5.1\,$Mpc$^2$). Additionally, $F606W$ ($V$-band)\textit{Hubble Space Telescope} (HST) Advanced Camera for Surveys imaging was obtained by Gray et al. (2009) covering $>90$\% of the COMBO-17 footprint, with gaps only at the outer edge (see geometric coverage in Figure~2 of Gray et al. 2009). 24~$\mu$m photometric coverage from \textit{Spitzer} (Bell et al. 2007), and X-ray data from \textit{XMM-Newton} (Gilmour et al. 2007) are also available. Hubble morphological types were derived by Wolf et al. (2009) from visual classification of the \textit{HST} images.}

Photometric redshifts (Wolf et al. 2003) and stellar masses (Borch et al. 2006) 
have been calculated from the COMBO-17 spectral energy distributions (SEDs). Galaxies have been
grouped into one of three SED categories (``Blue Cloud'', ``Dusty Red'', or ``Old Red'') by 
Wolf, Gray, \& Meisenheimer (2005) based on application of a red sequence cut suggested by 
Bell et al. (2004) for COMBO-17. ``Blue Cloud'' galaxies are the star-forming galaxies on the blue side of the cut.
``Old Red'' galaxies are dust-free red sequence galaxies with $E_{\rm B-V}<0.1$.  ``Dusty Red'' galaxies
are red sequence galaxies with $E_{\rm B-V}>0.1$. 

The SED type naming convention is somewhat misleading as ``Dusty Red'' galaxies do not contain 
more dust than ``Blue Cloud'' galaxies, but their ongoing star formation is confined to more 
dust obscured regions. Wolf, Gray \& Meinenheimer (2005) and Wolf et al. (2009) demonstrated the 
``Dusty Red'' galaxies are actively forming stars at rates $\sim4$ times lower than ``Blue Cloud'' 
galaxies at fixed mass.  
See Rodr\'iguez del Pino et al. (2017) for additional discussion of these SED types.

Our cluster dataset further includes redshifts derived from spectroscopic, prism, and narrow-band imaging 
techniques. These redshifts are discussed below in detail.

\subsection[]{Spectroscopic Redshifts}\label{specz}
Spectroscopic follow-up of targeted galaxies was carried out with the 2 degree Field (2dF) spectrograph on the 
Anglo-Australian Telescope and the Very Large Telescope (VLT) VIsible MultiObject Spectrograph (VIMOS) instrument. 

The 2dF observations (Gray et al. 2009) prioritised galaxies in the multi-cluster photometric redshift space having 
$R$-band magnitude $m_{\rm R}<20$ and ultimately yielded redshifts for 356 galaxies. The 2dF redshifts ($z_{\rm 2dF}$) 
were measured from the Ca H and K absorption features and cross-correlation of the spectra with templates.
{The VIMOS campaign (B\"osch et al. 2013) provides spectra for 200 $M_\star > 10^{8.5}\,M_\odot$  
galaxies having visually confirmed disk components in the $HST$ images, a star-forming SED, and an inclination 
angle $i>30$\deg.}  VIMOS Redshifts ($z_{\rm VIMOS}$) were measured from emission lines in 188 cases and absorption 
lines in the other 12 cases. 

Gray et al. (2009) report a redshift accuracy for 2dF of $\Delta_{\rm z} = 0.00149$. 
B{\"o}sch et al. (2013) find a VIMOS redshift accuracy of $\Delta_{\rm z} = 0.00019$, based on
16 galaxies included in more than one mask. For the 39 galaxies 
with redshifts from both surveys, the average of the 2dF and VIMOS redshifts is adopted.

\subsection[]{Prism Redshifts}\label{prismz}
Additional optical spectra were acquired with the Inamori Magellan Areal Camera and Spectrograph (IMACS) on the 
Baade I 6.5$\,$m telescope at Las 
Campanas following the prism spectroscopy method devised by the PRIsm MUlti-object Survey (PRIMUS, Coil et al. 2011) 
team. IMACS can observe $\sim2500$ objects simultaneously in an 0.18 deg$^2$ field of view, yielding more simultaneous 
spectra than traditional spectroscopic methods but at lower 
spectral resolving power ($R\approx 40$). The galaxies for which successful redshifts are observed with this 
approach are disproportionately red, massive, and have early Hubble types.

\begin{figure*}
\begin{center}
\scalebox{0.5}{\includegraphics[trim={2.5cm 0 2cm 0}]{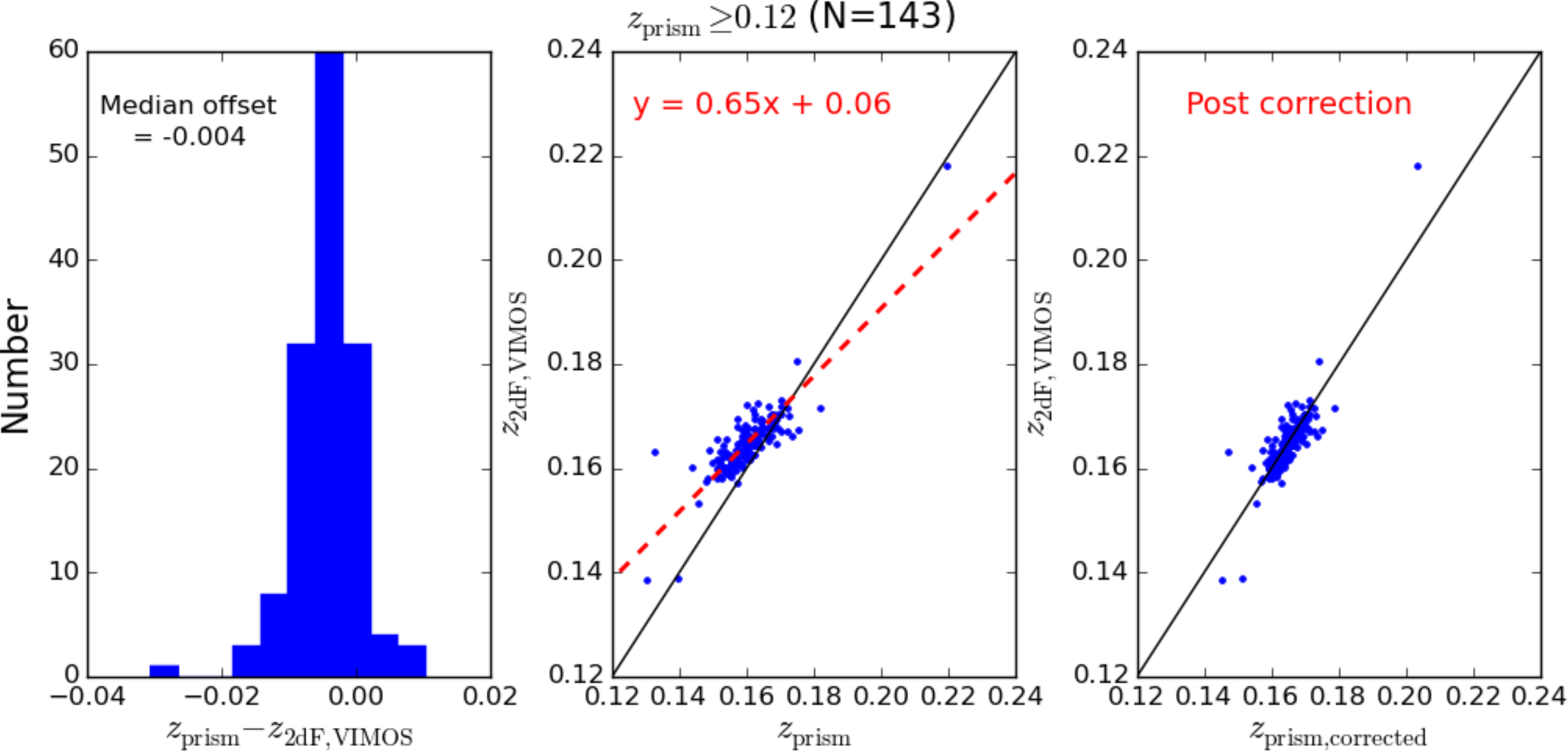}}
\caption{This figure compares the spectroscopic and prism redshifts for a subset of 143 galaxies having both
kinds of redshifts. The left panel shows the distribution of offsets between the spectroscopic and prism redshifts.
The middle panel shows that the offset is redshift dependent. The solid line is $y=x$.  The dashed line is a fit to 
the data.   In the right panel, we have used the fit from the middle panel to recalibrate the prism redshifts to 
better match the spectroscopic redshifts.
\label{primus}}
\end{center}
\end{figure*}

The accuracy of the prism redshifts ($z_{\rm prism}$) is evaluated with the subset of 143 sources with 
spectroscopic redshifts and $z_{\rm prism}\geq0.12$ (note, the prism redshifts catastrophically disagree with 
the spectroscopic redshifts for $z_{\rm prism}<0.12$). The left panel of Figure~\ref{primus} shows the prism 
redshifts are systematically lower than the spectroscopic redshifts by a median shift of 0.004. The middle panel 
indicates this offset has a redshift dependence that can be modelled as a straight line. Corrections derived from 
a linear model are applied to all prism redshifts so they better match the spectroscopic redshifts. The 
post-correction $1\sigma$ redshift scatter is $\sim0.0035$ for this subset.

\subsection[]{Emission-Line Redshifts From OMEGA}\label{omega}
OMEGA is based on a 90$\,$h ESO Large Programme survey with the Optical System for Imaging and low Resolution 
Integrated Spectroscopy (OSIRIS) instrument (Cepa et al. 2013) on the 10.4~m Gran Telescopio Canarias (GTC) 
telescope. The survey was designed to yield deep, spatially resolved emission-line images covering the H$\alpha$ and 
{\rm [N\textsc{ii}]} lines ($\lambda_{\rm obs} = 7595$--$7714\,$\AA) for galaxies in the STAGES multi-cluster. 
{OMEGA covers $\simeq658\,$arcmin$^2$, $\sim70$\% of the COMBO-17 footprint. Furthermore, 95\% of the OMEGA survey area is also covered by STAGES \textit{HST} imaging (see Figure~1 of Chies Santos et al.\ 2015 for a detailed description of the survey geometry).}

Low-resolution, $14\,$\AA\ full width at half maximum (FWHM) emission-line spectra containing 36 to 48 data points each 
were generated from the Fabry-P\'erot narrow-band imaging using two
different apertures. The radius of the aperture corresponds to 2.5 times the second central moment 
of the $F606W$ light distribution in the STAGES catalogue.  The diameter of the second, smaller 
aperture matches the 1.2 arcsec FWHM point spread function (PSF)
of the OSIRIS tunable filter imaging. The spectra have an absolute wavelength calibration
of $\sim1\,$\AA, equal to $\sim7\%$ of the instrumental resolution (Weinzirl et al. 2015). From the OMEGA 
spectra, redshifts, H$\alpha$ fluxes ($f_{\rm H\alpha}$), {\rm [N\textsc{ii}]} fluxes (if detected), and 
H$\alpha$ equivalent widths ($W_{\rm H\alpha}$) are measured via 
Bayesian Markov Chain Monte Carlo (MCMC) techniques. Chies-Santos et al. (2015) and 
Rodr\'iguez del Pino et al. (2017) provide details on the survey implementation and spectral fitting.  

The OMEGA sample of emission-line galaxies is defined with cuts in H$\alpha$ flux 
($f_{\rm H\alpha} \geq3\times 10^{-17}$~erg/s/cm$^2$), H$\alpha$ equivalent width ($W_{\rm H\alpha} \geq3\,$\AA), 
and photometric redshift. All sources in OMEGA that meet the cuts in flux and equivalent width, including 
contaminants and spurious detections, have emission-line redshifts ($z_{\rm OMEGA}$) in the narrow range 
[0.150, 0.176].  A cut in 
photometric redshift is necessary to isolate the true members of A901/2.  We choose
to restrict photometric redshift to the range [0.126, 0.200], which was derived by widening the [0.150, 0.176]
range by the 3$\sigma$ uncertainty (0.024) in photometric redshift for a source with $R$-band magnitude 20.5, the 
median brightness of OMEGA detections (see also Rodr\'iguez del Pino et al. 2017).

Galaxy spectra meeting the cuts in $f_{\rm H\alpha}$, $W_{\rm H\alpha}$, and photometric redshift are visually
inspected for quality. Each spectrum is reviewed independently by three 
classifiers. Spectra flagged two or more times for poor quality are removed from further consideration. After visual
classification, there are 439 large-aperture galaxy spectra in which H$\alpha$ is detected (321 of these also show
{\rm [N\textsc{ii}]}) and 360 PSF-aperture spectra containing both H$\alpha$ and {\rm [N\textsc{ii}]}.

Among galaxies with existing redshifts from 2dF/VIMOS, OMEGA detects all galaxies having rest-frame $B-V<0.7$.
More importantly, the majority (270) of the OMEGA detections did not previously have spectroscopic or prism redshifts.
The new redshifts are an important addition because OMEGA preferentially selects a population of blue star-forming 
galaxies that is under-represented in the 2dF and prism redshifts (note, the VIMOS redshift sample was selected
based on a star-forming SED type).

\begin{figure*}
\begin{center}
\scalebox{0.49}{\includegraphics[trim={0cm 3cm 0cm 3cm},clip]{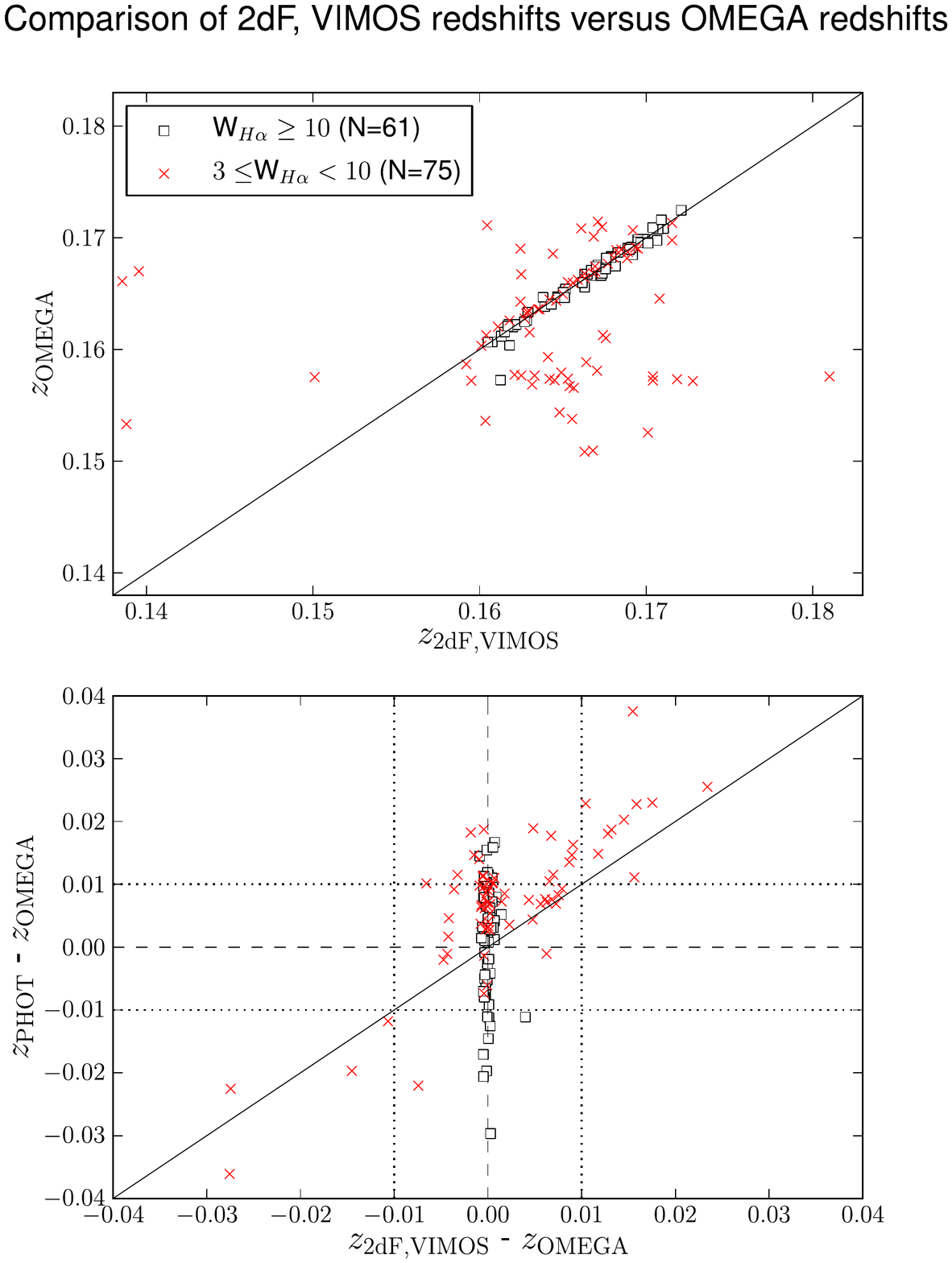}}
\caption{This figure compares  the OMEGA, spectroscopic, and photometric redshifts for the 136 sources in OMEGA that
also have spectroscopic redshifts from 2dF and/or VIMOS.  Top panel: Redshifts in OMEGA versus those from 2dF/VIMOS. 
Bottom panel: The difference of the OMEGA redshifts from the spectroscopic redshifts ($x$-axis) and the COMBO-17 
photometric redshifts ($y$-axis). Data points are coded by H$\alpha$ equivalent width ($W_{\rm H\alpha}$). 
The dashed and dotted lines show redshift offsets of 0 and $\pm0.01$, respectively. In both panels, the solid lines indicate equality.
The 1$\sigma$ standard deviation of the redshift bias between 2dF/VIMOS and OMEGA is $\sim0.0082$ for $3\leq W_{H\alpha}<10$ and $\sim0.00066$ for $W_{\rm H\alpha}\geq10$.  
\label{omega-specz}}
\end{center}
\end{figure*}

The top panel of Figure~\ref{omega-specz} compares the OMEGA redshifts to the spectroscopic redshifts for 
the 136 galaxies in OMEGA meeting the cuts in $f_{\rm H\alpha}$, $W_{\rm H\alpha}$, and photometric redshift. 
The agreement is almost perfect for sources with 
$W_{\rm H\alpha} \geq 10\,$\AA\ and some additional sources with lower equivalent width. There are 44 
sources of low equivalent width ($3\leq W_{\rm H\alpha}<10\,$\AA) having OMEGA redshifts offset by $>0.001$ from the
spectroscopic measurement, compared to just two at $W_{\rm H\alpha}\geq10\,$\AA. 

The bottom panel of 
Figure~\ref{omega-specz} compares the offsets of the OMEGA redshifts from the photometric and spectroscopic 
redshifts ($z_{\rm phot} - z_{\rm OMEGA}$ versus $z_{\rm 2dF,VIMOS} - z_{\rm OMEGA}$).  Data points on the
line of equality in the top panel are now on the dashed $x=0$ line. The problematic OMEGA
redshifts are generally offset from the spectroscopic and photometric redshifts in the same diagonal direction 
(i.e., they are smaller), suggesting the OMEGA redshifts are inaccurate because of spurious detections or 
inaccurate emission-line identification.

The $1\sigma$ standard deviation of the residuals between the spectroscopic and OMEGA redshifts is $\sim0.00820$ for $3\leq W_{\rm H\alpha}<10$ and $\sim0.00066$ for $W_{\rm H\alpha} \geq10$. 
Among the 439 galaxy spectra passing visual inspection, 
there are 120 with $3\leq W_{\rm H\alpha} <10$ and 319 with $W_{\rm H\alpha} \geq10$, 
so the average weighted error in the OMEGA redshifts is $\sim0.0027$.

\subsection[]{Definition of Parent and Redshift Samples}\label{sample}
\begin{figure*}
\begin{center}
\scalebox{0.45}{\includegraphics{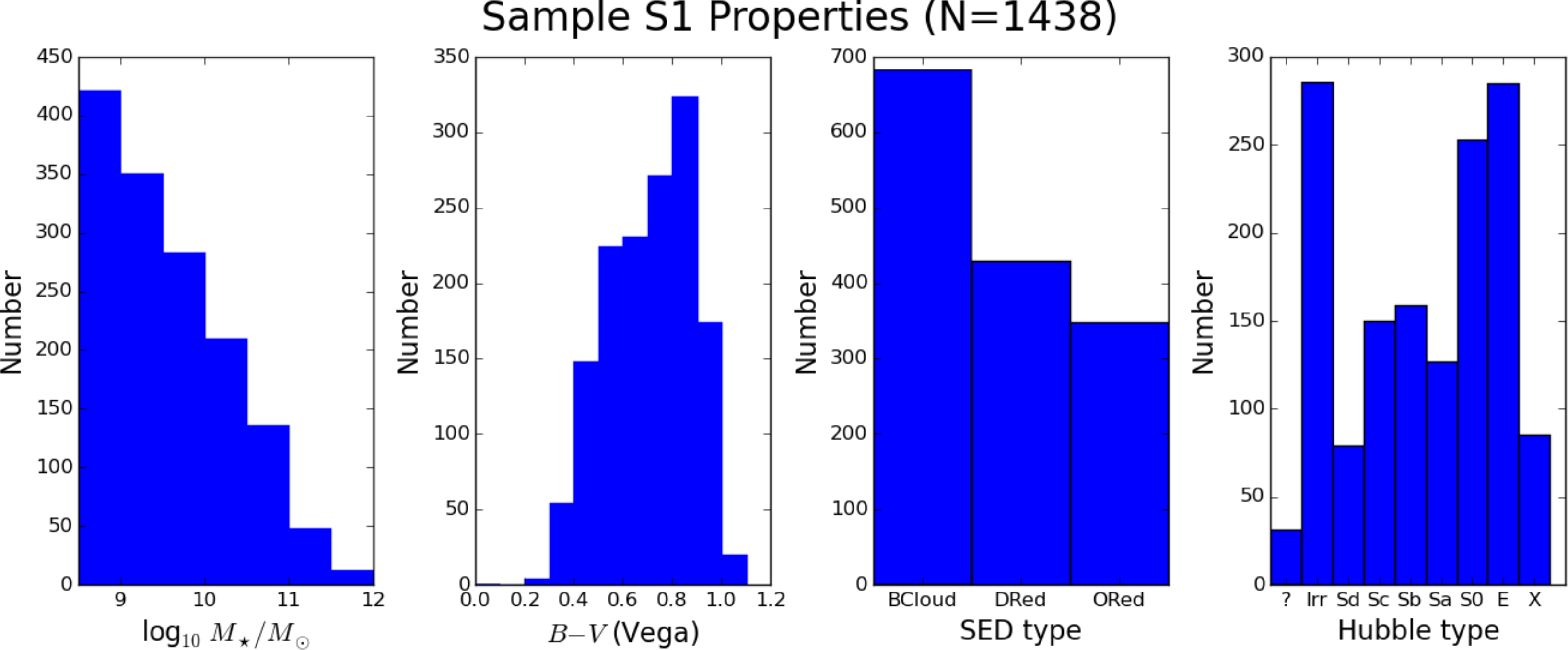}}
\caption{Distributions of stellar mass, colour, and Hubble type for sample S1 with $M_\star \geq 10^{8.5}$~$M_\odot$ 
and photometric redshift in the range [0.126, 0.200]. In the third panel from the left, the SED types are abbreviated 
as BCloud (``Blue Cloud''), DRed (``Dusty Red''), and ORed (``Old Red''). The fourth panel shows the Hubble types 
derived from the classifications of Wolf et al. (2009), where ``?'' means unclassifiable and ``X'' means resolved
but compact.
\label{mainsample-hist}}
\end{center}
\end{figure*}

The sample of galaxies used for the PPS analysis is presented here.
We first define a parent sample S1 of 1438 STAGES galaxies from cuts in stellar mass 
($M_\star \geq 10^{8.5}$~$M_\odot$) and photometric redshift [0.126, 0.200] to match
the OMEGA sample (Section~\ref{omega}). We further define selection criteria using STAGES catalogue flags
from Gray et al. (2009) to require galaxies be in the COMBO-17 footprint 
($\texttt{COMBO\_FLAG}\geq3$) and to have been detected in the $HST$ imaging ($\texttt{STAGES\_FLAG}\geq2$). 
Most ($\sim87\%$) of S1 galaxies are ``cluster member'' galaxies having STAGES catalogue flag
$\texttt{COMBO\_FLAG}\geq4$.  This selects all galaxies according to the definition of 
``cluster'' in Figure 13 of Gray et al. (2009). 
Figure~\ref{mainsample-hist} shows histogram distributions for stellar mass, colour, SED type, morphology of S1. 
The mass function of S1 does not turn over above $M_\star \geq 10^{8.5}$~$M_\odot$. Gray et al. (2009) show STAGES
is $>90\%$ complete for $R<23.5$, which corresponds to $M_\star \geq 10^9 M_\odot$ (Maltby et al. 2010). 
The completeness of our sample should be somewhat less than $90\%$ given our choice of
mass cut.

Some of the OMEGA detections (Section~\ref{omega}) are excluded from S1 because they either fall below the stellar mass 
limit or they have $\texttt{COMBO\_FLAG}=2$. S1 contains 397/439 OMEGA galaxies with large-aperture spectra and 
329/360 galaxies with PSF-aperture spectra.
 
\begin{figure*}
\begin{center}
\scalebox{0.50}{\includegraphics{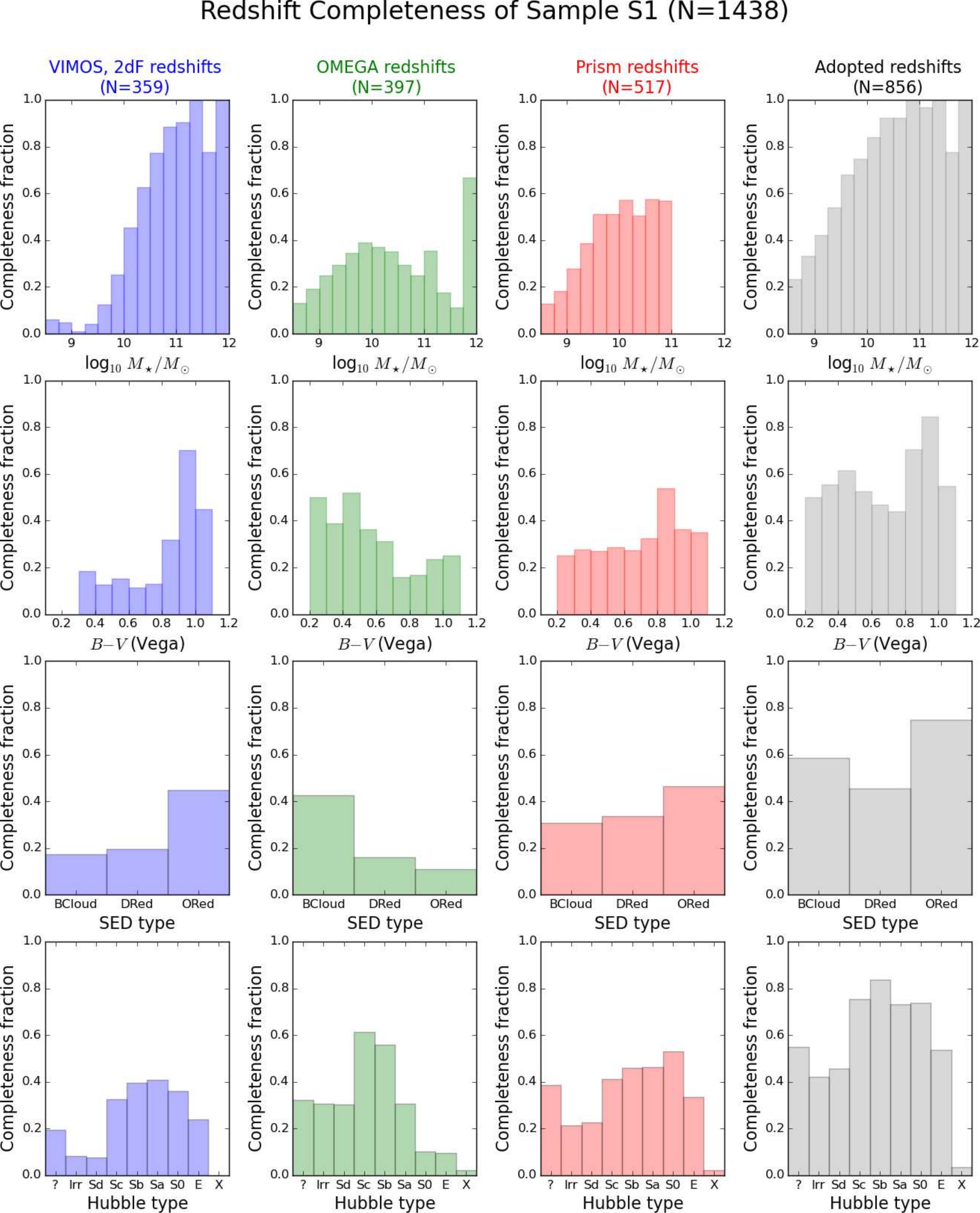}}
\caption{This figure shows the redshift completeness fraction of the mass and photometric redshift 
selected sample, S1, for stellar mass, colour, SED type, and morphology.  The SED types and morphologies were
classified by Wolf et al. (2009). Columns 1-3 represent the 2dF/VIMOS redshifts, OMEGA redshifts, and prism 
redshifts, respectively.  Column 4 represents the best available redshifts adopted in descending order
(Column 1, Column 2, Column 3) of spectral resolution. The SED and Hubble type labels
are as in Figure~\ref{mainsample-hist}.
\label{zcompleteness}}
\end{center}
\end{figure*}

Figure~\ref{zcompleteness} shows the redshift completeness fractions for sample S1 in terms of stellar 
mass, colour, SED type, and morphology for the different redshift measures. The inherent selection bias of
each redshift technique is apparent in columns~1--3.  The spectroscopic and prism
redshift samples have relatively few blue, star-forming galaxies, which is the dominant population detected by
OMEGA.

Sample S2 is constructed from the 856 galaxies with redshift measurements. The best available redshift, in 
descending order of spectral resolution (2dF/VIMOS, OMEGA, and prism spectroscopy) is chosen for each galaxy.
Column~4 of Figure~\ref{zcompleteness} represents sample S2.  The redshift completeness in stellar mass is $>50\%$ 
for $M_\star >10^{9.5}$~$M_\odot$, is $>40\%$ across colour, and is $>50\%$ for Hubble types E to Sc.  
Including galaxies uniquely detected by OMEGA in this analysis increases the number of redshifts by $\sim47\%$ over 
what just VIMOS/2dF and prism spectroscopy provide alone.

Sample S2, while limited in completeness, benefits from combining the sensitivities of 
each of the three redshift surveys it was built from.  The result is a sample covering 
a much better range of galaxy properties (colour, SED type, morphology).
We will use this sample throughout the rest of the paper. {Sample S2 contains 856 galaxies, of which 359 have redshifts based on 
2dF/VIMOS spectroscopy, 273 based on OMEGA data, and 224 based on prism observations. }

\section[]{Phase Space}\label{sphasespace}
{For the purposes of the dynamical analysis in this paper we subdivide the field into regions surrounding four main subclusters.  Here we are guided by the luminosity-weighted maps of photometrically selected cluster members and peaks in the gravitational lensing mass maps (Gray et al.\ 2002;  Heymans et al.\ 2008).}

For sample S2, the positions of galaxies in PPS are calculated relative to the nearest subcluster
in A901/2. Specifically, we use the subcluster-centric radius normalised by
the subcluster virial radius ($R_{\rm p}/R_{\rm 200}$) and the peculiar line-of-sight rest-frame velocity
normalised by the subcluster galaxy velocity dispersion ($\Delta V_{\rm los}/\sigma_{\rm scl}$).
We adopt the subcluster centres and $R_{\rm 200}$ 
from the ``one-halo'' models fitted by Heymans et al. (2008). We use the definition 
\begin{equation}
\Delta V_{\rm los} = \frac{c(z-z_{\rm scl})}{(1+z_{\rm scl})}, 
\end{equation}
where $c$ is the speed of light and $z_{\rm scl}$ is the subcluster redshift.

Mean subcluster redshifts $z_{\rm scl}$ 
and velocity dispersions $\sigma_{\rm scl}$ are calculated with the biweight estimators of location and 
scale (Beers et al. 1990) following the procedure employed by B\"osch et al. (2013) in their previous calculations 
for A901/2. These calculations warrant repeating because there are now significantly more galaxy redshifts (856) 
than the 200 VIMOS redshifts used in B\"osch et al. (2013).  Galaxies are initially assumed to be
members of the nearest subcluster in terms of projected distance. Subsequent iterations use $3\sigma$ 
clipping to filter outliers until convergence where all remaining galaxies fall into the $3\sigma$ velocity-dispersion 
interval of their subcluster. 

{At this point it is important to consider how the relatively large uncertainties of the OMEGA and prism redshifts 
($\delta z\sim0.003$ or $\delta v \sim 900\,$km$\,$s$^{-1}$) may affect the velocity dispersion determination and our subsequent analysis. Reassuringly, the values we obtain agree very well with the ones published by B\"osch et al. (2013): three out of the four $\sigma_{\rm scl}$ values agree  within $1\sigma$ error, and the fourth is $\sim1.8\sigma$ away. Moreover, we detect no bias between our measurements and theirs. For additional peace of mind, we have also calculated $\sigma_{\rm scl}$ using the more accurate 2dF and VIMOS redshifts only.  
We find that the subcluster redshifts $z_{\rm scl}$ change by less than $0.001$ in all cases, and $\sigma_{\rm scl}$ changes by less than 10\% in 3 out of 4 cases (the exception is the SW group, where the uncertainty in the 2dF/VIMOS value is very large due to small number statistics, and the different values agree well within the errors). We are therefore satisfied that using the complete set of redshifts does not bias the measured redshifts or velocity dispersions. Because the formal errors are smaller when using the larger galaxy sample, we have carried our analysis using the $z_{\rm scl}$ and $\sigma_{\rm scl}$ values we calculated. However, our conclusions would not change had we used the compatible values estimated by B\"osch et al. (2013).}

{Table~\ref{tab:table1} lists the resulting subcluster redshifts and velocity dispersions along with the adopted centres and dark halo parameters ($R_{\rm 200}$, $M_{\rm 200}$) from Heymans et al.\ (2008). 
Note that the derived physical parameters of the clusters (cluster centres, $R_{\rm 200}$, $M_{\rm 200}$) were not determined from the velocity dispersion measurements but from the lensing work of Heymans et al.\ (2008). Velocity dispersions are affected by the dynamical state of the structures, and determining subcluster masses and radii from $\sigma_{\rm scl}$ is very uncertain and potentially biased for non-virialised clusters and/or in the presence of substructure.}

\begin{table*}
  \centering
  \caption{Subcluster redshifts, velocity dispersions, and dark halo parameters in A901/2}
  \label{tab:table1}
  \begin{tabular}{lcccccc}
	& & & & & & \\
Subcluster & RA (Degrees) & DEC (Degrees) & $z_{\rm scl}$ & $\sigma_{\rm scl}$ (km/s) & ${R_{\rm 200}}$ (Mpc) & $M_{\rm 200}$ (10$^{14}$ $M_\odot$) \\
\hline
A901a &  149.1099   & $-$9.9561  &    0.1631 &  878$^{+17}_{-27}$ &  0.84$^{+0.06}_{-0.07}$ & 1.3$^{+0.3}_{-0.3}$\\
& & & & & & \\
A901b &  148.9889   & $-$9.9841  &    0.1641 &  937$^{+15}_{-22}$ &  0.83$^{+0.06}_{-0.07}$ & 1.3$^{+0.3}_{-0.3}$ \\
& & & & & & \\
A902 &   149.1424   & $-$10.1666  &    0.1656 &  808$^{+13}_{-20}$ &  0.56$^{+0.08}_{-0.09}$ & 0.4$^{+0.2}_{-0.2}$ \\
& & & & & & \\
SW Group & 148.9101 & $-$10.1719  &  0.1693 &  585$^{+25}_{-37}$ &  0.63$^{+0.07}_{-0.08}$ & 0.6$^{+0.2}_{-0.2}$ \\
\hline
\multicolumn{7}{l}{\textbf{Table notes:} }\\
\multicolumn{7}{l}{The cluster redshifts and velocity dispersions ($z_{\rm scl}$, $\sigma_{\rm scl}$) are determined in Section~\ref{sphasespace}.} \\
\multicolumn{7}{l}{The cluster centres and dark halo parameters ($R_{\rm 200}$, $M_{\rm 200}$) come from Heymans et al. (2008), {and are not}}\\
\multicolumn{7}{l}{{based on the values of the subclusters velocity dispersions (see text for details).}} \\
\end{tabular}
\end{table*}

\begin{figure*}
\begin{center}
\scalebox{0.50}{\includegraphics{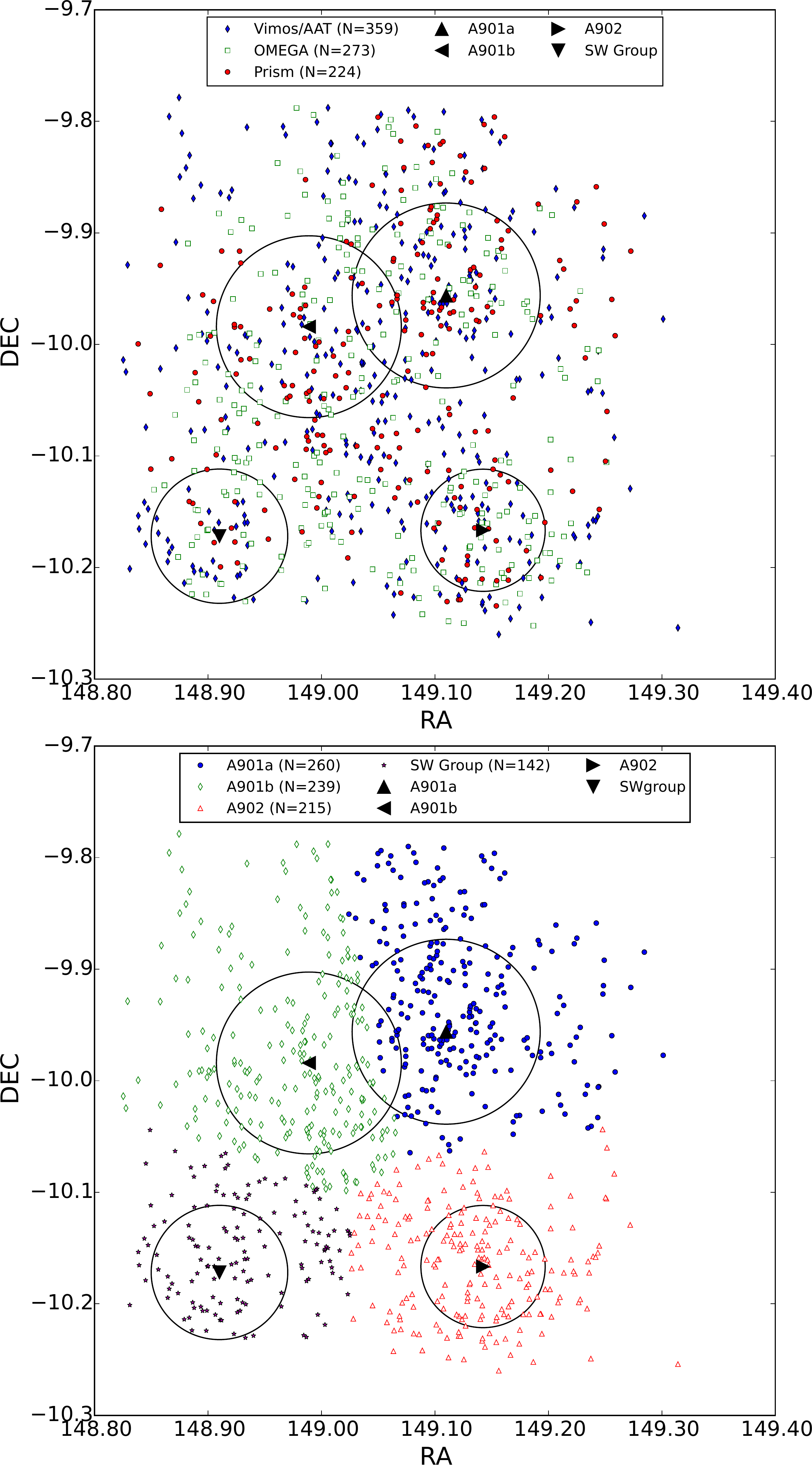}}
\caption{The top panel shows the spatial distribution of the 856 galaxies in sample S2 
coded by the source of the adopted redshifts. The four subclusters are separately 
labelled, and their spheres of influence ($R_{\rm 200}$) are shown as circles. In the bottom
panel, galaxies are colour coded by the nearest cluster centre.
\label{radec}}
\end{center}
\end{figure*}

Figure~\ref{radec} shows the spatial distribution of sample S2 in relation to the subcluster centres. In the top
panel, galaxies are coded by the adopted redshift source. In the bottom panel, galaxies are coded by the nearest
cluster centre. Phase space parameters ($R_{\rm p}/R_{\rm 200}$, $\Delta V_{\rm los}/\sigma_{\rm scl}$) are 
calculated relative to the halo parameters of the nearest cluster,
a strategy that maximises the number of galaxies in the central virialised region. Potential concerns with this
strategy are the effects of overlapping $R_{\rm 200}$, cluster centring errors, and contamination between clusters.
The $R_{\rm 200}$ overlap for
A901a and A901b, causing 20 galaxies to fall inside both $R_{\rm 200}$ radii. Our cluster membership scheme equally
divides the overlapping galaxies between A901a and A901b. The close proximity of A901a and 
A901b should not seriously bias the phase space analysis because their halo parameters are similar;
their $R_{\rm 200}$ agree to within 0.01 Mpc, and their $\sigma_{\rm scl}$ differ only by 10\% 
(Table~\ref{tab:table1}).

We use the PPS to distinguish cluster galaxies in the virialised region from infalling galaxies with the aim of 
constraining which environmental processes are relevant in A901/2.
Following Mahajan, Mamon \& Raychaudhury (2011) and Jaff\'e et al.
(2015), we use triangular regions to represent the inner virialised region. 
Our fiducial boundary (B1), $R_{\rm p}/R_{\rm 200} \leq 1.2$ and 
$|\Delta V_{\rm los}/\sigma_{\rm scl}| \leq 1.5-1.5/1.2 \times R_{\rm p}/R_{\rm 200}$, is used by Jaff\'e et al. (2015) in 
the study of a cluster that is comparable in mass ($7.7\times 10^{14}$~$M_\odot$) to A901a and at a comparable redshift 
($z=0.2$). 
The second boundary (B2), ${R_{\rm p}/R_{\rm 200}} \leq 0.5$ and
$|\Delta V_{\rm los}/\sigma_{\rm scl}| \leq 2.0-2.0/0.5 \times R_{\rm p}/R_{\rm 200}$, 
extends to higher $\Delta V_{\rm los}/\sigma_{\rm scl}$. While B2 is set somewhat 
arbitrarily, it defines a virialised region 45\% smaller in area than B1; its purpose is
to provide an estimate of how sensitive the phase-space analysis is on the definition of
the virialised region. 

\begin{figure*}
\begin{center}
\scalebox{0.70}{\includegraphics[trim={4cm 0cm 0cm 0cm},clip]{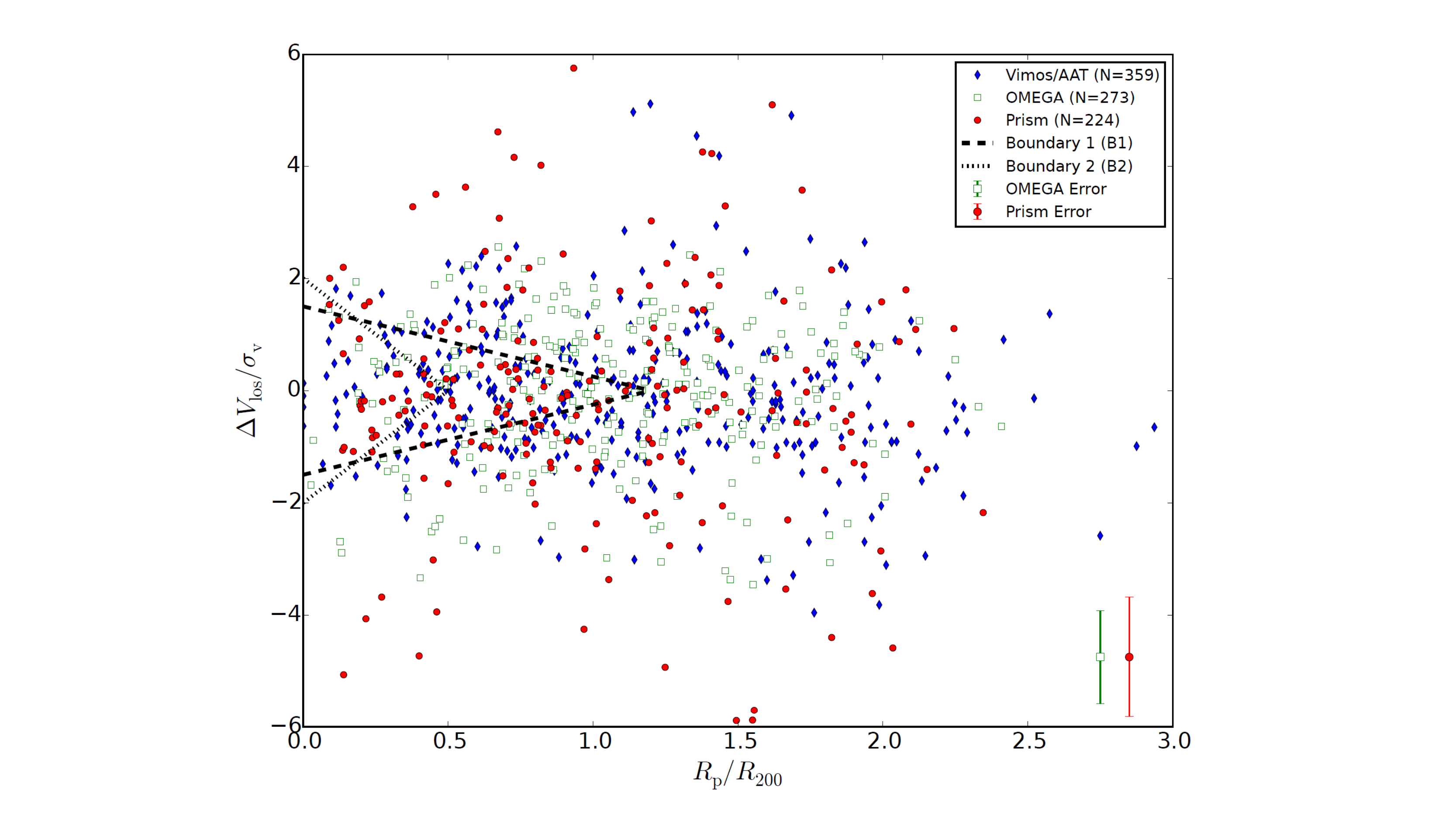}}
\caption{The projected phase-space of the 856 galaxies in sample S2 are shown, coded by the source of the adopted 
redshifts. The four subclusters are stacked together.  The dashed and dotted lines denote the two different boundaries 
used for the virialised region of phase-space. The dashed line (Boundary 1) is defined by 
$R_{\rm p}/R_{\rm 200} \leq 1.2$ and 
$|\Delta V_{\rm los}/\sigma_{\rm scl}| \leq 1.5-1.5/1.2 \times R_{\rm p}/R_{\rm 200}$. The dotted line (Boundary 2) 
is defined by $R_{\rm p}/R_{\rm 200} \leq 0.5$ and
$|\Delta V_{\rm los}/\sigma_{\rm scl}| \leq 2.0-2.0/0.5 \times R_{\rm p}/R_{\rm 200}$.
There are 210 galaxies inside Boundary 1 and 70 galaxies inside Boundary 2. The average errors for the OMEGA and prism 
$|\Delta V_{\rm los}/\sigma_{\rm scl}|$ appear in the lower right, and they reflect the redshift biases relative
to the spectroscopic redshifts.  
\label{phasespace}}
\end{center}
\end{figure*}

Figure~\ref{phasespace} is a PPS diagram with these boundaries overlaid, stacking together the four subclusters in 
A901/2. B1 and B2 are represented with dashed and dotted lines, respectively. Galaxies are again coded by adopted 
redshift source. B1 encloses 210 galaxies while B2 encloses 70 galaxies.  

An important caveat to keep in mind
is the potential effect of spatial incompleteness on the stacked PPS diagram.  Sources of spatial incompleteness 
include, for example, the finite OMEGA footprint and 2dF fibre collisions in crowded virialised regions.
While Figure~\ref{radec} shows all clusters are well populated within $R_{\rm 200}$, spatial incompleteness is 
evident beyond $R_{\rm 200}$, particularly for A902 and the SW Group. {Given the narrow redshift range covered by the system, possible incompleteness in the redshift sampling are not expected to be a significant concern. }

{Visual inspection of Figure~\ref{phasespace} reveals that the PPS diagram of the A901/902 system does not show the clear ``trumpet shape'' expected for spherically collapsing systems (e.g., Regos \& Geller 1989; Diaferio \& Geller 1997). This could be due (at least in part) to the relatively large redshift uncertainties affecting a significant fraction of the galaxy sample, but it could also be the consequence of the complex nature and dynamical state of the whole system and each individual subsystem. If each individual subcluster could be considered as not disturbed, we should be able to see a ``trumpet shape'' when stacking the four distributions, provided we are able to cleanly separate members of each structure (which is clearly not the case due to complex spatial and redshift projection effects). Because it is highly likely that the four subclusters are interacting and possibly merging with each other, it is hard to believe that they are individually unperturbed. Redshift uncertainties would, no doubt, contribute to blurring any putative underlying ``trumpet shape'', but we think the complexity of the system, which compounds dynamical and projection effects, is the dominant reason why we do not see such a shape. The distribution on this diagram of the galaxies with 2dF/VIMOS spectroscopy (for which the redshift uncertainties are much smaller) is not significantly different from that of the galaxies with prism or OMEGA redshifts. This suggests that even if we had \textit{perfect} redshifts for the whole sample, a clear ``trumpet shape'' would not arise. }

\section[]{Environmental Effects Across Projected Phase Space}\label{dependence}

\subsection{Imaging Properties}\label{photoprop}
For sample S2, we look for differences in imaging-derived galaxy properties (stellar mass, rest-frame colour, 
structure, Hubble type, and SED type) between the virialised and non-virialised regions of the PPS.  The stellar 
masses, colours, and SED types were derived from the COMBO-17 photometry, and Hubble types were derived from
the $HST$ images (Section~\ref{data}).  
The structural parameters include galaxy-wide S\'ersic (1963)
indices and half-light radii measured from the $HST$ images by
Gray et al. (2009). S\'ersic index is interesting because it broadly distinguishes 
dynamically cold, disky stellar structures from dynamically hot ones (Andredakis, Peletier, \& Balcells 1995, 
de Jong 1996, Khosroshahi, Wadadekar, \& Kembhavi 2000, and more recently e.g., Weinzirl et al. 2011, 2014).

\begin{figure*}
\begin{center}
\scalebox{0.48}{\includegraphics[trim={0cm 0cm 0cm 0cm}]{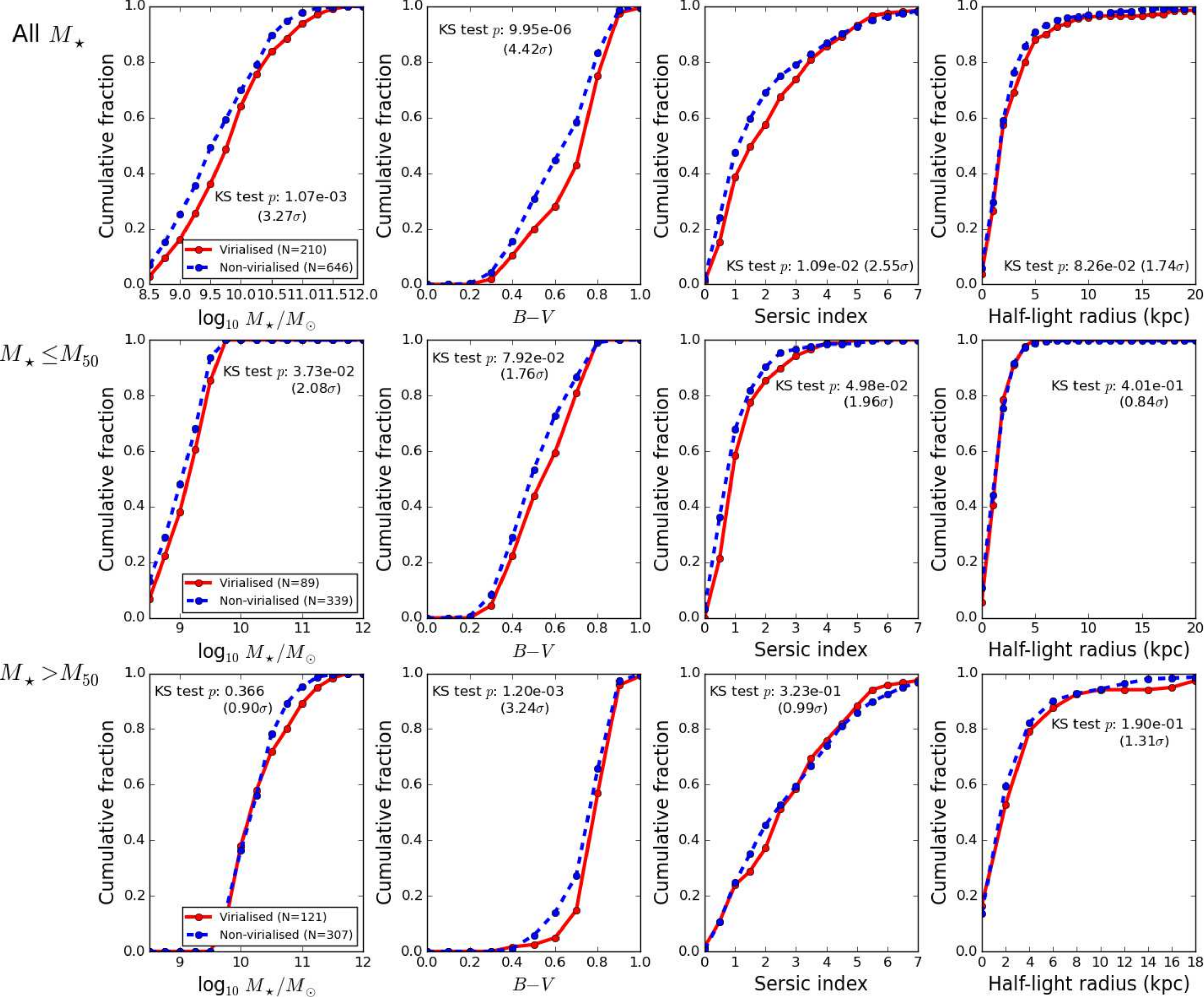}}
\caption{
Stellar mass, colour, S\'ersic index, and half-radius are compared in the virialised and non-virialised 
regions of the A901/2 system. The virialised region is defined using fiducial boundary B1 (Section~\ref{sphasespace}, Figure~\ref{phasespace}). Panel annotations convey the $p$-value significances of Kolmogorov-Smirnov (KS) tests on the distributions in each panel. 
The top row includes galaxies of all stellar masses. The middle and bottom rows represent subsamples
partitioned around the 50th percentile in stellar mass ($M_{50} = 10^{9.85}~M_\odot$).
Similar results are found when substituting boundary B1 for B2.
\label{env-ksmass50}}
\end{center}
\end{figure*}

The top row of Figure~\ref{env-ksmass50} shows the cumulative distributions of stellar mass, rest-frame $B-V$ colour, S\'ersic index, 
and half-light radius inside and outside the virialised regions, as defined by our fiducial boundary B1 
(Section~\ref{sphasespace}, Figure~\ref{phasespace}). Kolmogorov-Smirnov (KS) tests are used to compare 
the statistical significance of apparent disparities across phase-space. For stellar mass, there is a 3.3$\sigma$ 
difference ($p=0.0011$) such that the galaxies in the virialised region are more massive. There is a significantly 
stronger (4.4$\sigma$, $p=1.0\times 10 ^{-5}$) disparity in colour implying these central 
galaxies are redder overall. S\'ersic index and half-light radius show less significant 
differences. 
At the 2.6$\sigma$-level, virialised-region galaxies tend to have higher S\'ersic indices (i.e., they are less 
disky) than galaxies outside the virialised region.  Virialised-region galaxies may also be slightly 
larger in half-light radius (1.7$\sigma$).

Galaxy colour (e.g., Bell et al. 2004), galaxy size (e.g., Lange et al. 2015), and S\'ersic index 
(e.g., Graham \& Guzm{\'a}n 2003, Kormendy et al. 2009) all scale with galaxy luminosity (i.e., stellar mass).
Since galaxies in the virialised region are more massive (the median stellar mass is 
$10^{10.03}$~$M_\odot$ in the virialised region versus $10^{9.76}$~$M_\odot$ in the non-virialised region), it is 
important to consider if it is mass, rather than environmental effects, driving the differences in the other three 
properties. 

The middle and bottom rows of Figure~\ref{env-ksmass50} repeat the KS tests of the virialised and non-virialised regions for the sample 
after subdivision into bins above and below the 50th percentile in stellar mass ($M_{50}=10^{9.85}$~$M_\odot$) within sample S2. 
For the lower-mass bin ($M_\star \leq M_{50}$), $B-V$ and S\'ersic index show differences 
across the PPS at marginal significance that can probably
be accounted for by stellar mass.
For the higher-mass bin ($M_\star > M_{50}$), there is no evidence
for changes in stellar mass, S\'ersic index, or half-light radius.
This distributions of $B-V$, however, differ very significantly ($>3\sigma$).

\begin{figure*}
\begin{center}
\scalebox{0.64}{\includegraphics[trim={0cm 0cm 0cm 0cm}]{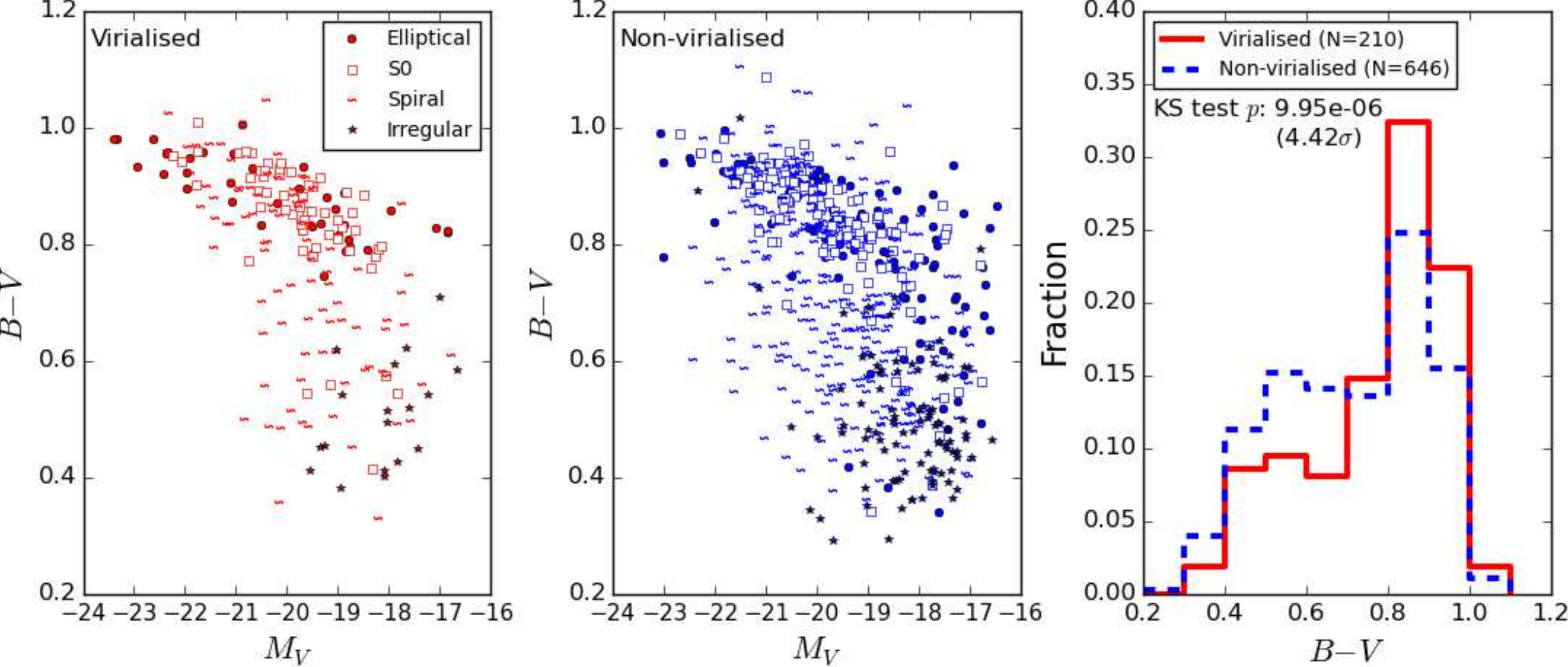}}
\caption{The left and middle panels show the rest-frame $B-V$ colour versus absolute $V$ magnitude diagram for the 
virialised and non-virialised regions. The right panel shows histograms of colour in the two phase-space bins.
Like Figure~\ref{env-ksmass50}, the virialised region is defined using fiducial boundary B1.
\label{env-bv}}
\end{center}
\end{figure*}

The change in $B-V$ but not stellar mass in the $M_\star > M_{50}$ row of Figure~\ref{env-ksmass50} suggests the 
4.4$\sigma$ result in the top row for $B-V$ is not driven entirely by mass.  Figure~\ref{env-bv} shows
the colour magnitude diagrams and distributions of $B-V$ colour in the two phase-space bins. The colour-magnitude diagrams 
(left and middle panels) show proportionately 
fewer galaxies with $B-V<0.7$ at fixed $M_{\rm V}$ in the virialised region (left) compared to the non-virialised region (middle). 
Across all luminosities, the non-virialised region contains a factor of 1.6
more galaxies with $B-V<0.7$ than the virialised region. Noteworthy in the non-virialised region
is a population of spirals with $M_{\rm V} \leq -21$ (i.e., $M_\star \geq 10^{10}$~$M_\odot$) that 
is absent in the virialised region.
Thus, the strong colour difference in Figure~\ref{env-ksmass50} arises from a deficit of blue ($B-V <0.7$), 
high-mass ($M_\star>10^{9.85}$~$M_\odot$) galaxies (mostly spirals) in the 
virialised region. 

\begin{figure*}
\begin{center}
\scalebox{0.55}{\includegraphics[trim={0cm 0cm 0cm 0cm}]{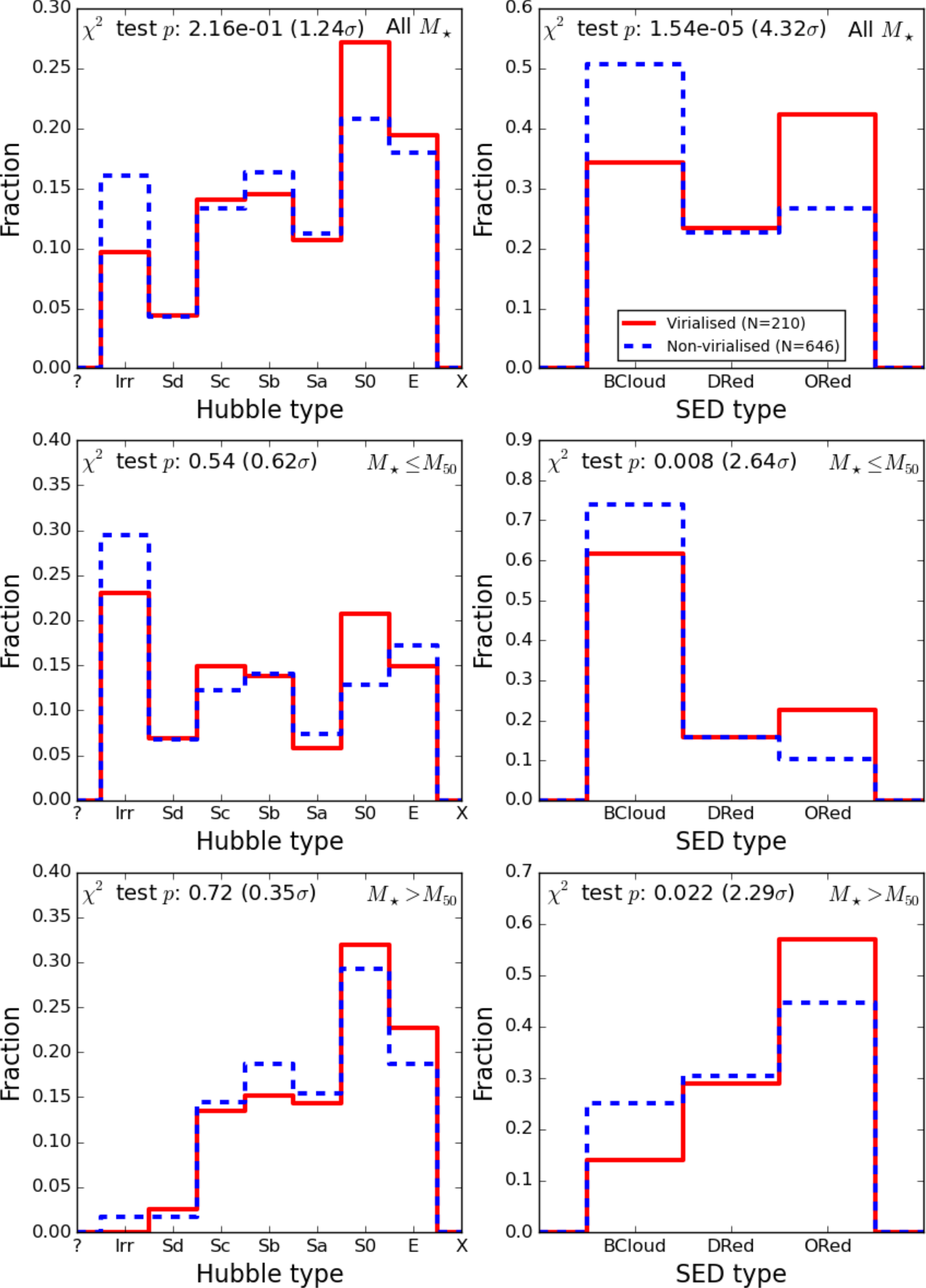}}
\caption{This figure addresses the disparity in morphology (left column) and SED type (right column) across the two 
phase-space bins. The top row makes the comparison using all stellar masses. The middle and bottom rows repeat the 
analysis for the lower-mass bin ($M_\star \leq M_{50}$) and higher-mass bin ($M_\star > M_{50}$), 
respectively. Like Figure~\ref{env-ksmass50}, the virialised region is defined using fiducial boundary B1.
\label{env-chi2}}
\end{center}
\end{figure*}

A similar analysis is conducted to assess the influence of stellar mass on morphology and SED type.  The left column 
of Figure~\ref{env-chi2} shows the histograms of Hubble type in three different stellar mass bins (all masses,
$M_\star \leq M_{50}$, $M_\star > M_{50}$).  Visual inspection of
the histogram for all masses shows irregular galaxies are $\sim30\%$ less common in the virialised region and that 
S0s are $\sim25\%$ more common.  A $\chi^2$ test (rather than KS, since these are discrete quantities)
for Hubble types E to Irr, however, gives a $p$-value of 0.216 (1.2$\sigma$), indicating 
these differences are not statistically meaningful. The middle and bottom panels 
show Hubble type has even less variation across the PPS once isolated into low and high mass bins.

The right column of Figure~\ref{env-chi2} addresses SED type. Across all stellar masses, there is a 4.3$\sigma$ 
difference between ``Blue Cloud'' and ``Old Red'' SED types, such that there are $\sim50\%$ ($\sim62\%$) more 
(fewer) ``Blue Cloud'' (``Old Red'') galaxies in the non-virialised region. This result is stronger than the 
result for stellar mass (3.3$\sigma$) and comparable to the one for rest-frame $B-V$ (4.4$\sigma$) in 
Figure~\ref{env-ksmass50}. In the $M_\star \leq M_{50}$ mass bin, this difference drops to 2.6$\sigma$ but is 
still more significant than the change in stellar mass across the PPS in the low-mass bin (2.1$\sigma$). In the $M_\star > M_{50}$ mass bin, the difference for SED type is 2.3$\sigma$, 
much higher than the 0.90$\sigma$ found for stellar mass. Thus, SED type evolves across the PPS similarly to 
rest-frame $B-V$.

The KS and $\chi^2$ tests presented in this section have focused on the fiducial boundary B1.  The test statistics 
are summarised in columns $3-5$ of Table~\ref{tab:table3}.  Similar and qualitatively consistent results are
found for alternate boundary B2, which defines a more conservative virialised region (Section~\ref{sphasespace}). 
Column~6 lists the test statistics when applying instead boundary B2 without binning by 
stellar mass.

\begin{table*}
\small
  \centering
  \caption{Changes in galaxy properties across stellar mass, PPS}
  \label{tab:table3}
  \begin{tabular}{lccccc}
\hline
Property &Method & PPS, B1$^a$ & PPS, B1$^a$ ($M_\star \leq M_{50}$)& PPS, B1$^a$ ($M_\star > M_{50}$)& PPS, B2$^b$ \\
(1) & (2) & (3) & (4) & (5) & (6) \\
\hline
Stellar mass      & KS test       & 1.07$\times10^{-3}$ (3.27$\sigma$)& 3.73$\times10^{-2}$ (2.08$\sigma$)& 3.66$\times10^{-1}$ (0.90$\sigma$)& 1.17$\times10^{-2}$ (2.52$\sigma$)\\
\hline
Colour ($B-V$)    & KS test       & 9.95$\times10^{-6}$ (4.42$\sigma$)& 7.92$\times10^{-2}$ (1.76$\sigma$)& 1.20$\times10^{-3}$ (3.24$\sigma$)&9.33$\times10^{-7}$ (4.91$\sigma$)\\
\hline
S\'ersic index    & KS test       & 1.09$\times10^{-2}$ (2.55$\sigma$)& 4.98$\times10^{-2}$ (1.96$\sigma$)& 3.23$\times10^{-1}$ (0.99$\sigma$)&4.71$\times10^{-3}$ (2.83$\sigma$)\\
\hline
Half-light radius & KS test       & 8.26$\times10^{-2}$ (1.74$\sigma$)& 4.01$\times10^{-1}$ (0.84$\sigma$)& 1.90$\times10^{-1}$ (1.31$\sigma$)&3.71$\times10^{-1}$ (0.89$\sigma$)\\
\hline
Hubble type       & $\chi^2$ test & 2.16$\times10^{-1}$ (1.24$\sigma$)& 5.38$\times10^{-1}$ (0.62$\sigma$)& 7.25$\times10^{-1}$ (0.35$\sigma$)& 3.14$\times10^{-2}$ (2.15$\sigma$)\\
\hline
SED type          & $\chi^2$ test & 1.54$\times10^{-5}$ (4.32$\sigma$)& 8.34$\times10^{-3}$ (2.64$\sigma$)& 2.20$\times10^{-2}$ (2.29$\sigma$)& 1.39$\times10^{-8}$ (5.67$\sigma$)\\
\hline
\end{tabular}
    \begin{tablenotes}
      \scriptsize 
    \item \textbf{Table notes:}\\
    $M_{50}=10^{9.85}$~$M_\odot$ is the 50th-percentile in stellar mass.\\
       \item $^{a}$Tests the change in galaxy property inside and outside the virialised region in phase-space defined by boundary B1, $R_{\rm p}$/$R_{200} \leq 1.2$ and 
     $|\Delta V_{\rm los}$/$\sigma_{\rm scl}| \leq 1.5-1.5/1.2 \times R_{\rm p}/R_{200}$.
    \item $^{b}$Tests the change in galaxy property inside and outside a smaller virialised
    region in phase-space defined by boundary B2, $R_{\rm p}$/$R_{200} \leq 0.5$ and 
    $|\Delta V_{\rm los}$/$\sigma_{\rm scl}| \leq 2.0-2.0/0.5 \times R_{\rm p}/R_{200}$.
    \end{tablenotes}
\end{table*}

\subsection{Emission-Line Derived Properties}\label{specprop}
In this section, we explore how emission-line properties derived from the OMEGA spectra
(gas-phase metallicity, star formation, and AGN emission) vary across the PPS.

\subsubsection[]{Star-Formation and AGN Activity}\label{sfVagn}
\begin{figure*}
\begin{center}
\scalebox{0.75}{\includegraphics[trim={0cm 6cm 0cm 9cm},clip]{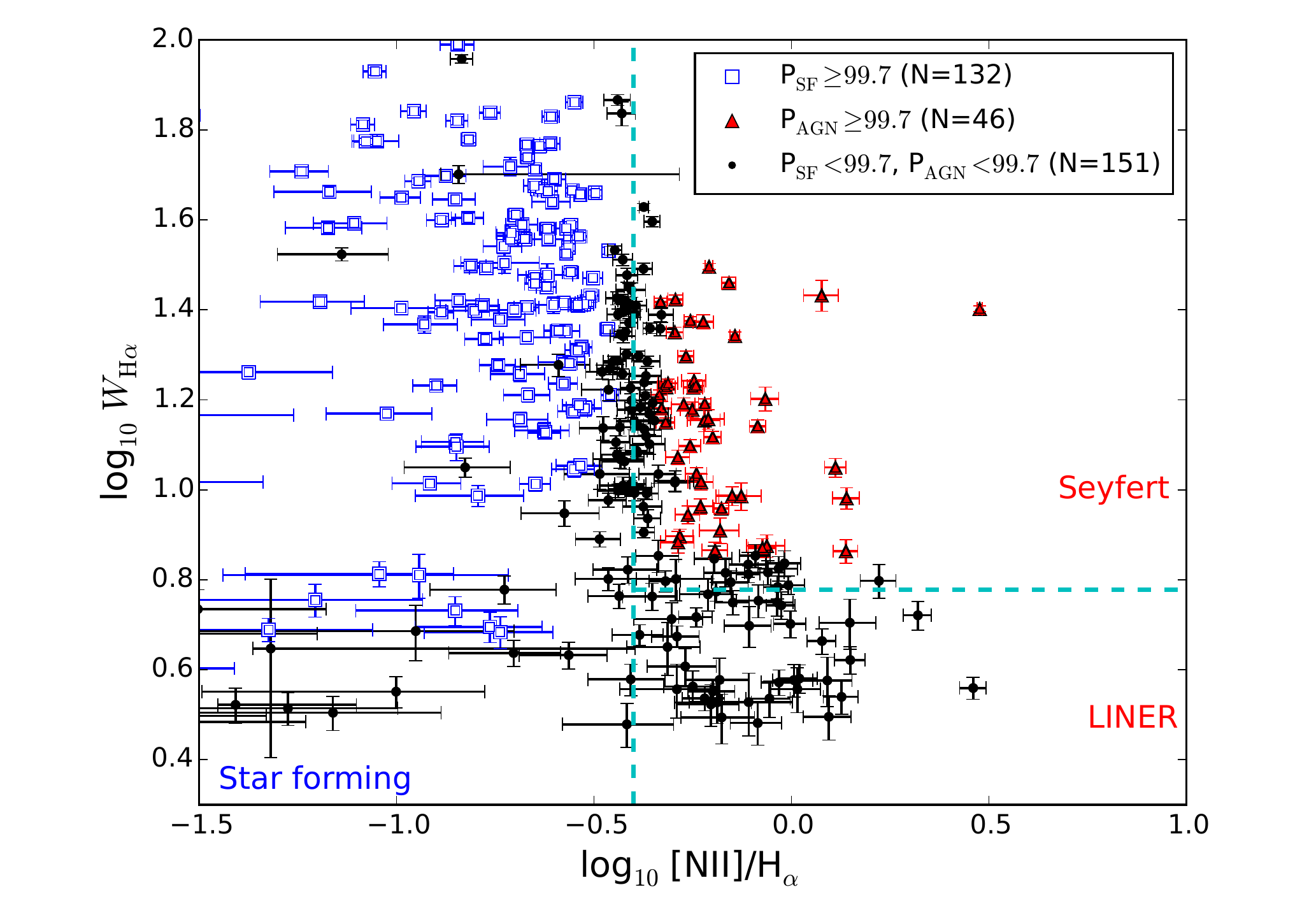}}
\caption{This figure shows the WHAN diagram ({\rm [N\textsc{ii}]}/H$\alpha$ gas-phase metallicity versus 
H$\alpha$ equivalent width, $W_{\rm H\alpha}$), introduced by Cid Fernandes et al. (2010, 2011), for 
the 329 OMEGA galaxies in sample S2 (Section~\ref{sample}) whose PSF-aperture spectra 
have both the H$\alpha$ and {\rm [N\textsc{ii}]} lines within the wavelengths probed.  
Galaxies that are dominated by star formation or AGN activity at probability $\geq 99.7\%$ are 
distinguished. See Section~\ref{sfVagn} for details.
\label{env-whan}}
\end{center}
\end{figure*}

Cid Fernandes et al. (2010, 2011) demonstrate the utility of using the WHAN diagram 
({\rm [N\textsc{ii}]}/H$\alpha$ versus $W_{\rm H\alpha}$) to discriminate star formation and black hole 
activity as sources of gas ionisation. Figure~\ref{env-whan} shows the WHAN diagram, assuming the standard 
boundaries from Cid Fernandes et al. (2010), for the subset of 329 OMEGA galaxies in sample S2 whose
PSF-aperture spectra 
have both the H$\alpha$ and {\rm [N\textsc{ii}]} within the wavelengths probed. 
There are 198/329 
($\sim60.2\%$) galaxies on the ``Star forming'' side of the diagram, 93/329 ($\sim28.3\%$) in the ``Seyfert'' area, 
and 38/329 ($\sim11.6\%$) in the ``LINER'' area. 

The significance of the ``Star forming'' versus ``AGN'' classifications in the WHAN diagram can be quantified
using the probability distributions for {\rm [N\textsc{ii}]}/H$\alpha$ and $W_{\rm H\alpha}$ arising from our 
Bayesian spectral fitting (Section~\ref{omega}). The probability that a galaxy is dominated by star 
formation or AGN emission is calculated based on its position relative to the boundaries in Figure~\ref{env-whan}. Galaxies with 
$\geq 99.7\%$ ($\geq3\sigma$) probability of being in the star-forming or AGN categories are distinguished in 
Figure~\ref{env-whan}. There are 132 bona-fide such star-forming galaxies and 46 such AGN-dominated galaxies.

Table~\ref{tab:table2} lists the number counts of $\geq 3\sigma$ star-forming and AGN galaxies in different regions
of the PPS, as well as in the two stellar mass bins $M_\star \leq M_{50}$ and $M_\star > M_{50}$.
The majority of star-forming (105/132 for B1 and 125/132 for B2) and AGN galaxies (29/46 for B1 and 39/46 for B2) 
are outside the virialised region. Most star-forming galaxies (108/132) are in the low 
mass bin, while the majority (31/46) of AGN are in the high mass bin. 

A $\chi^2$ test for the numbers of star-forming and AGN galaxies across the PPS using the B1 boundary
gives a $p$-value of 0.025 ($2.2\sigma$). The same test for B2 gives
a similar $p$-value, 0.031 (2.2$\sigma$). The $\chi^2$ test across mass bins gives a highly
significant $p$-value of 5.0$\times10^{-10}$ (6.2$\sigma$).  Change in the numbers of star-forming and AGN 
galaxies across phase-space is quite weak compared to the change in number counts due to stellar mass.

\begin{table}
  \centering
  \caption{Number of $\geq3\sigma$ AGN and star-forming galaxies}
  \label{tab:table2}
  \begin{tabular}{lcc}
& & \\
Condition & $\geq 3 \sigma$ Star forming & $\geq 3 \sigma$ AGN \\
\hline
Virialised region (B1$^a$) & 27  & 17 \\
Non-virialised region (B1$^a)$ & 105 & 29 \\
\hline
Virialised region (B2$^b$) & 7  & 7 \\
Non-virialised region (B2$^b$) & 125 & 39 \\
\hline
$M_\star \leq M_{50}^c$ & 108 & 15 \\
$M_\star > M_{50}^c$ & 24  & 31 \\
\hline
\end{tabular}
    \begin{tablenotes}
      \scriptsize 
    \item \textbf{Table notes:}\\
    \item $^{a}$The virialised region in phase-space is defined by boundary B1, $R_{\rm p}$/$R_{200} \leq 1.2$ and 
     $|\Delta V_{\rm los}$/$\sigma_{\rm scl}| \leq 1.5-1.5/1.2 \times R_{\rm p}/R_{200}$.\\
    \item $^{b}$
    The virialised region in phase-space defined by boundary B2, $R_{\rm p}$/$R_{200} \leq 0.5$ and 
    $|\Delta V_{\rm los}$/$\sigma_{\rm scl}| \leq 2.0-2.0/0.5 \times R_{\rm p}/R_{200}$.\\
    \item $^{c}$ $M_{50}=10^{9.85}$~$M_\odot$ is the 50th-percentile in stellar mass.\\
    \end{tablenotes}

\end{table}

\subsubsection{Metallicity in Star-Forming Galaxies}\label{metallicity}

The {\rm [N\textsc{ii}]}/H$\alpha$ ratio has been shown (e.g., Denicol{\'o}, Terlevich, \& Terlevich 2002;
Pettini \& Pagel 2004) 
to be a robust calibrator of gas-phase metallicity [$\rm O/H$] for star-forming galaxies. 
The relation of {\rm [N\textsc{ii}]}/H$\alpha$ to metallicity
is monotonic, and the {\rm [N\textsc{ii}]} and H$\alpha$ lines can be measured in our moderate-resolution OMEGA
spectra. The downside of {\rm [N\textsc{ii}]}/H$\alpha$ is that its sensitivity to ionisation variations 
(e.g., Nagao, Maiolino, \& Marconi 2006) means it is only suitable for measuring integrated galaxy abundances.
Here, we consider the galaxy-wide {\rm [N\textsc{ii}]}/H$\alpha$ metallicities of the $\geq 3 \sigma$ 
star-forming galaxies \textit{identified from}
PSF-aperture OMEGA spectra in Section~\ref{sfVagn} and Figure~\ref{env-whan} (details of the photometry are
in  Rodr\'iguez del Pino et al. 2017).

\begin{figure*}
\begin{center}
\scalebox{0.50}{\includegraphics[trim={0.6cm 0 2cm 0}]{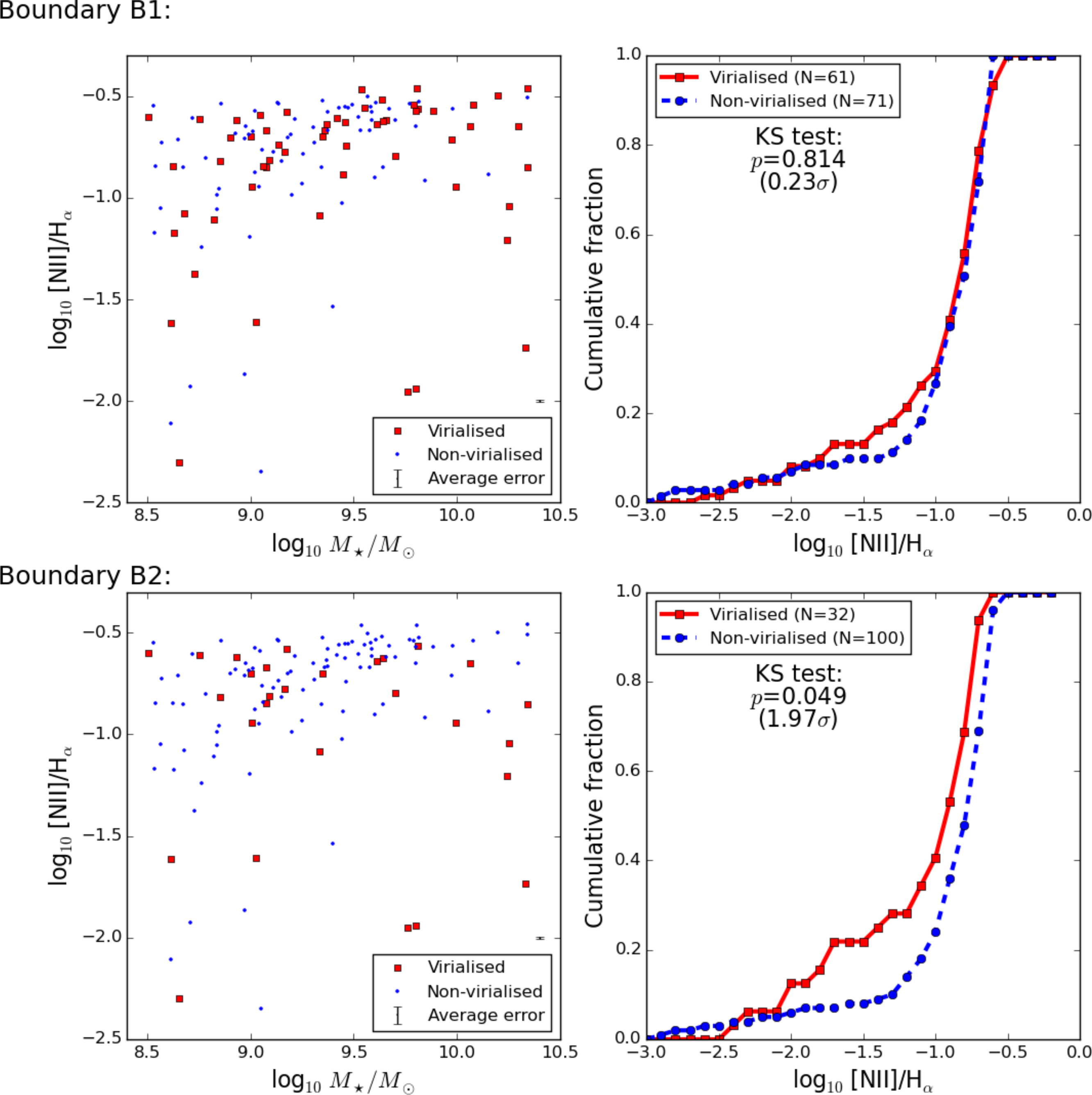}}
\caption{This figure explores potential change in gas-phase metallicity among the 132 galaxies whose PSF-aperture 
OMEGA spectra are dominated by star formation at the $\geq99.7\%$ ($\geq3\sigma$) level (see 
Section~\ref{sfVagn} for details).  
Top row: The left panel shows the {\rm [N\textsc{ii}]}/H$\alpha$ ratio versus stellar mass, colour-coded 
by position in the PPS (with respect to the PPS boundary B1).
The right panel shows the corresponding cumulative distributions and results of a KS test. Bottom row: Same as 
top row, except for the PPS boundary B2.
\label{env-n2ha}}
\end{center}
\end{figure*}

The top left panel of Figure~\ref{env-n2ha} shows the {\rm [N\textsc{ii}]}/H$\alpha$ ratio versus stellar mass, 
colour-coded by position in the PPS, with respect to boundary B1. There is a weak trend in 
{\rm [N\textsc{ii}]}/H$\alpha$ across stellar mass present for both PPS bins. Moreover, the full range of
line ratio is present in the virialised and non-virialised regions.
The top right panel of Figure~\ref{env-n2ha}
compares the cumulative distributions for the two PPS bins.
A KS test gives a $p$-value 0.81 (0.23$\sigma$). There is no evidence that 
the distribution of {\rm [N\textsc{ii}]}/H$\alpha$ differs between the two regions.

The bottom row of Figure~\ref{env-n2ha} repeats the same analysis for the more central 
PPS boundary B2. The results appear to be more significant for B2
($p=0.049$, $2.0\sigma$), and the cumulative line ratio distributions suggest there are more
galaxies with low metallicities ($\rm {[N\textsc{ii}]}/H\alpha <-1$) in the virialised region.  
However, the significance of the KS test is still only marginal with just 32 bona-fide
star-forming galaxies in the more conservative virialised region. We conclude that in our dataset we do not detect any significant difference in metallicity as traced by {\rm [N\textsc{ii}]}/H$\alpha$ across the PPS.

\subsubsection[]{H$\alpha$ Specific Star Formation Rate Deficit}\label{ssfr}

Using H$\alpha$ star-formation rates ($SFR_{\rm H\alpha}$) derived from the large-aperture OMEGA spectra of
galaxies in sample S2 (Section~\ref{sample}), we look for a change in the way specific $SFR_{\rm H\alpha}$
(or $SSFR_{\rm H\alpha}$) scales with stellar mass in the \textit{intra-cluster} environment. As in 
Rodr\'iguez del Pino et al. (2017), we convert H$\alpha$ luminosity to star-formation rate
using the Kennicutt (1998) relation with an additional mass-dependent dust attenuation correction estimated 
from Figure 6 of Brinchmann et al. (2004). The conversion is 
\begin{equation}
\frac{SFR_{H\alpha}}{(M_\odot \rm{yr}^{-1})} = 7.9 \times 10^{\left [-42+0.4\times A_{H\alpha} \left( M_\star \right) \right]} \frac {L_{H\alpha}} {(\rm{erg s}^{-1})}
\end{equation}
where $A_{H\alpha} \left( M_\star \right)$ is the dust attenuation term.

\begin{figure*}
\begin{center}
\scalebox{0.65}{\includegraphics[trim={0cm 7cm 0cm 8cm},clip]{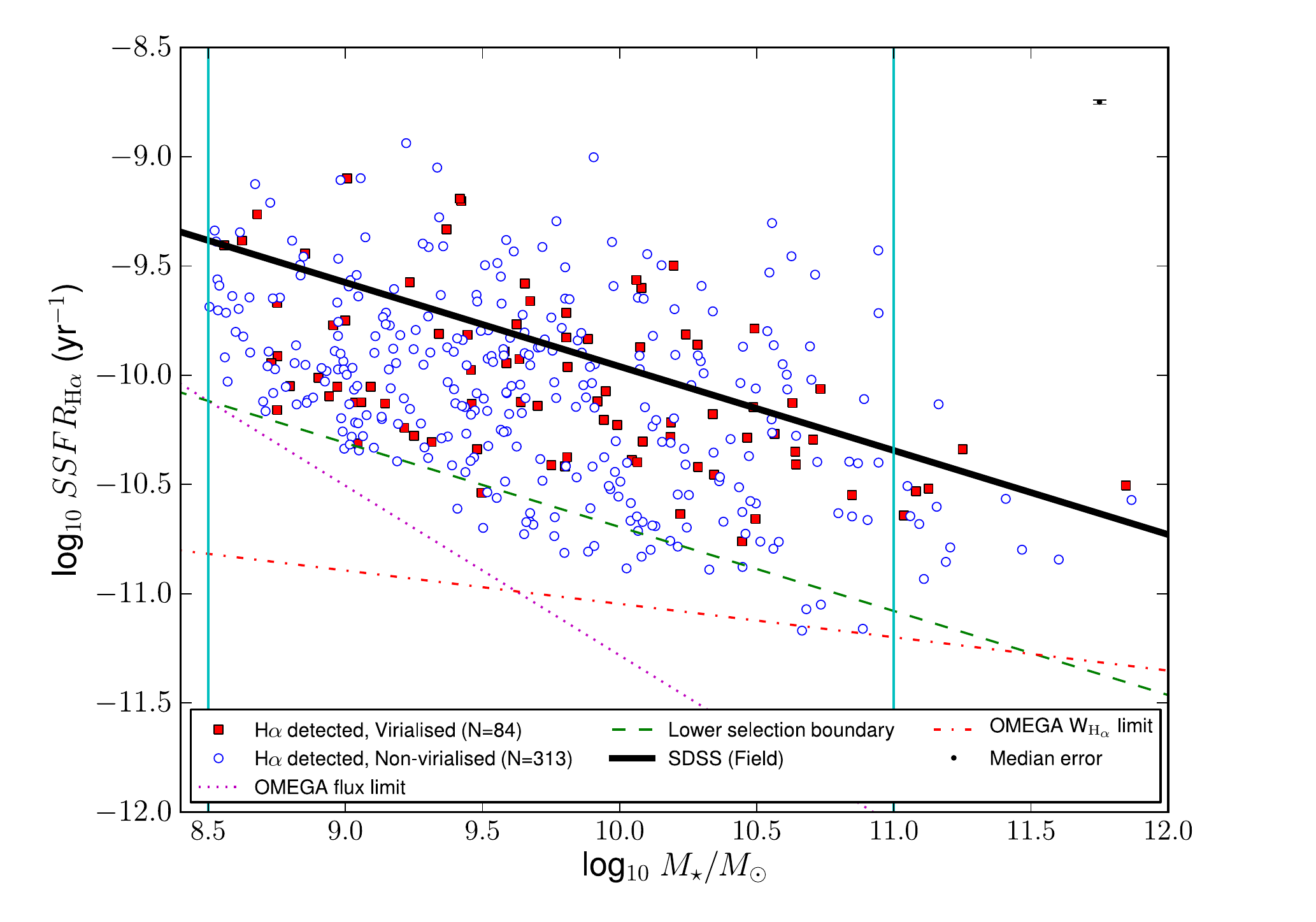}}
\caption{The $SSFR_{\rm H\alpha}$--$M_\star$ relation (Section~\ref{ssfr}) is compared for the field and the 397 
galaxies in sample S2 (Section~\ref{sample}) with large-aperture spectra from OMEGA (Section~\ref{omega}).  
The black line is the $SSFR_{\rm H_\alpha}$--$M_\star$ relation from SDSS (Abazajian et al. 2009).  The dashed line 
is a boundary parallel to the SDSS relation that intersects the H$\alpha$ flux limit 
($3\times 10^{-17}$ erg/s/cm$^2$) in OMEGA (dotted line).  The dash-dotted line represents the $3\,$\AA\  
H$\alpha$ equivalent width ($W_{\rm H_\alpha}$) detection limit. Sample S3 includes the 351 galaxies above the dashed 
line and between the vertical lines. The points are colour-coded by galaxy position in 
the PPS, as defined by fiducial 
boundary B1. The error bar in the upper right corner shows the median error on $SSFR_{\rm H_\alpha}$. 
See also Rodr\'iguez del Pino et al. (2017) for a related diagram.
\label{ssfr-def}}
\end{center}
\end{figure*}

We define $SSFR_{\rm H\alpha}$ deficit as the offset relative to the 
$SSFR_{\rm H\alpha}$--$M_\star$ relation for field galaxies in SDSS (Abazajian et al. 2009); at a given mass, a 
galaxy with a lower $SSFR_{\rm H\alpha}$ than the SDSS relation has a positive $SSFR_{\rm H\alpha}$ deficit. 
Figure~\ref{ssfr-def} shows the $SSFR_{\rm H\alpha}$--$M_\star$ relation for the field 
(black line) and for our sample (points) of 397 galaxies from S2 with large-aperture OMEGA spectra. 

To compare the star-formation deficit of the galaxies in different PPS regions we require an unbiased sample that is not affected by H$\alpha$ flux and equivalent-with selection effects. Following the detailed analysis of Rodr\'iguez del Pino et al. (2017), for this comparison we use sample S3, which comprises the 351 OMEGA galaxies lying above the dashed line shown in Figure~\ref{ssfr-def}. This line is parallel to the $SSFR_{\rm H\alpha}$--$M_\star$ relation in the field and defines a region where the H$\alpha$-detected galaxies are above the flux and equivalent-width detection limits of the survey (see Figure~\ref{ssfr-def} and its caption for more details). While we are only concerned here with the relative offset from the field line shown in Figure~\ref{ssfr-def} for galaxies in different regions of the PPS, it is clear that there is an overall reduction in $SSFR$ for the complete OMEGA sample relative to the field.  This is a result discussed in detail by Rodr\'iguez del Pino et al. (2017) for a similar sample drawn from the OMEGA survey.  

\begin{figure*}
\begin{center}
\scalebox{0.80}{\includegraphics[trim={0cm 5cm 0cm 6cm},clip]{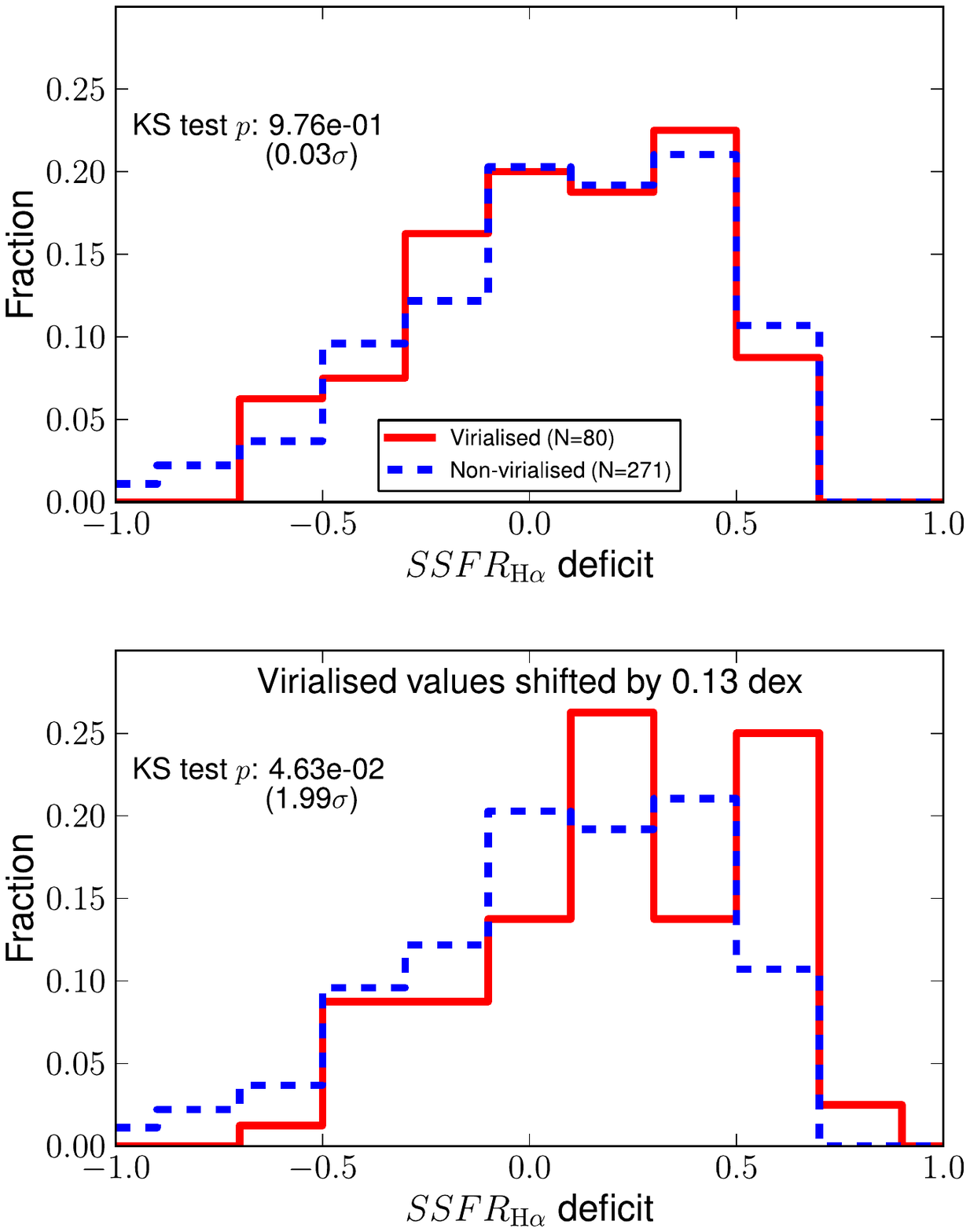}}
\caption{The top panel shows $SSFR$ deficit histograms for sample S3 (Figure~\ref{ssfr-def}) in the virialised and 
non-virialised regions
(as defined by boundary B1).  A KS test comparing $SSFR$ deficit in the two PPS regions shows no significant 
difference ($p=0.98$).  The bottom panel indicates a distinction would be detectable at the $\geq 2 \sigma$ level 
if the star formation deficits in the virialised region were increased by $\sim0.13$ dex or more.
\label{ssfr-deficit-test}}
\end{center}
\end{figure*}

The top panel of Figure~\ref{ssfr-deficit-test} shows histograms of the $SSFR_{\rm H\alpha}$ deficit in the virialised 
and non-virialised regions as defined by boundary B1.  A KS test shows no probable difference ($p=0.98$, $0.03\sigma$)
between these distributions.  In order to estimate the upper limit on the difference in average 
$SSFR_{\rm H\alpha}$ between the two regions that our data are able to rule out, we repeat the KS test for a range of relative offsets.
The bottom panel of Figure~\ref{ssfr-deficit-test} indicates a difference would be 
detectable at the $\geq 2 \sigma$ ($p \leq 0.046$) level if the $SSFR_{H\alpha}$ deficits in the 
virialised region were increased by $\sim0.13$~dex or more.  Using boundary B2 instead shows that an even larger
shift (0.18~dex or more) is required for a $\geq 2\sigma$ result. Thus, we rule out changes in $SSFR_{\rm H\alpha}$ deficit across PPS diagram regions larger than 40\%--50\%.  

\section[]{Discussion}\label{sdiscuss}

\subsection{Disentangling the Effects of Stellar Mass and Environment}\label{discuss1}
Table~\ref{tab:table3} summarises the significance of changes in the imaging-derived properties 
across PPS and stellar mass. 
An important point to note is that the galaxy properties considered here
($B-V$, S\'ersic index, half-light radius, Hubble type, and SED type) have trends with
stellar mass that are much more significant ($p<10^{-26}$) than their trends across
virialised and non-virialised PPS bins (Columns 3 through 6). 
The effect of mass on these galaxy properties is far stronger than the environmental effects revealed by the PPS.  

Columns~4 and~5 of Table~\ref{tab:table3} summarise the results of repeating the PPS analysis 
(Figures~\ref{env-ksmass50} and \ref{env-chi2}) for boundary B1 after binning by stellar mass.  The 
significance of the $p$-values is reduced, in part because of the reduced sample sizes. However, as noted 
in Section~\ref{photoprop}, the variations of colour and SED type across the PPS at $M_\star > M_{50}$
are more significant ($3.2\sigma$ and $2.3\sigma$) than the change in stellar mass ($0.90\sigma$). The changes in 
colour and SED in the higher mass bin cannot be explained by stellar mass alone and must be a result of environment. {To test to what extent mass differences within each bin could affect this conclusion, we have repeated the analysis using narrower mass bins (4 bins instead of 2, which is as far as we can go given the sample size). We are still able to detect environmentally driven colour changes at the $\sim2.5$--$3\sigma$ level in 3 out of the 4 bins, and the results are inconclusive in the other one. This new test supports the conclusions that environmentally driven colour changes are present at fix stellar mass. }

Column~6 further shows the comparison using the alternate, more conservative PPS boundary B2. The $p$-values 
for boundary B2 are qualitatively similar to those for boundary B1, with the significance for 
changes in colour, S\'ersic index, and especially SED type being higher in Column 6 due to
the more conservative virialised region boundary. The disparity in
colour (SED type) across all masses is $2.40\sigma$ ($3.2\sigma$) more significant than the change in 
stellar mass. Conducting the PPS analysis with
boundary B2 after binning by stellar mass likewise yields results consistent with Columns~4 and~5.
That the same qualitative results are seen with the alternative PPS boundary indicates the phase-space 
analysis is robust to the choice of the virialised region boundary.

Table~\ref{tab:table3} shows that S\'ersic index has a somewhat more significant, although still somewhat marginal,
change across PPS than Hubble type. This may seem puzzling at first sight since S\'ersic index usually correlates with Hubble type, albeit 
with large scatter. We must not forget that the visually-determined Hubble type reflects multiple
criteria, including bulge-to-disk ratio and the presence and tightness/smoothness of the spiral arms, while S\'ersic index reflects only the relative dominance of the bulge or the disk component in a galaxy (e.g., Kormendy \& Bender 2012). Since S\'ersic index and morphology are not the same, finding differences in environmental behaviour should not be surprising. Furthermore, considering that the significance of the environmental differences detected in the S\'ersic index distribution is small (always $<3\sigma$), and given the indirect connection between S\'ersic index and visual morphology, this apparent discrepancy is not that surprising. 

Thus, we find in A901/902 that colour appears to be changing independently of morphology and structure, with
the ``Dusty Red'' spirals being the prime examples of this effect in A901/902. {This result is in agreement
with previous studies concluding that the environment can drive changes in colour without changing 
morphology (e.g., Bamford et al.\ 2009; Skibba et al.\ 2009; Masters et al.\ 2010).}

\subsection{Properties of Star-Forming Galaxies}\label{discuss2}
We find multiple indications that the star-formation properties of star-forming galaxies do not differ significantly across the PPS in A901/2.
First, the proportion of $\geq3\sigma$ star-forming galaxies (Section~\ref{sfVagn}, Table~\ref{tab:table2}) changes 
only at the $\sim2.2\sigma$ level across the PPS, less significant than the change with stellar mass ($\sim6.2\sigma$).  
Likewise, there is little evidence ($<2\sigma$) for changes in the {{\rm{ [N\textsc{ii}]}/H$\alpha$}} 
metallicity (Section~\ref{metallicity}) and $SSFR_{\rm H\alpha}$ 
deficit (Section~\ref{ssfr}). 

Wolf, Gray \& Meinenheimer (2005) and Wolf et al. (2009) find ``Dusty Red'' galaxies in A901/2 make up a significant fraction of the star-forming galaxies. 
These ``Dusty Red'' galaxies constitute over half of star-forming galaxies at stellar masses in the range $10^{10}$--$10^{11}$~$M_\odot$, 
and they form stars $\sim4$ times slower at fixed mass than blue galaxies in A901/2.  The fact that many
star-forming galaxies are red explains why rest-frame colours and
SED types vary more strongly (Table~\ref{tab:table3}) across the PPS than star formation properties.  

Wolf et al. (2009) report an overall reduction in $SSFR_{\rm UV+IR}$ of star-forming galaxies 
relative to the field, and a subtle change in 
$SFR_{\rm UV+IR}$ across local density. Rodr\'iguez del Pino et al. (2017) report at fixed mass a reduction in 
$SSFR_{\rm H\alpha}$ of star-forming galaxies relative to the field. We see this offset too (Figure~\ref{ssfr-def}), but we do not see a difference in $SSFR_{\rm H\alpha}$ deficit within bins of the PPS (Figure~\ref{ssfr-deficit-test}).
A reduction in $SSFR_{\rm H\alpha}$ relative to the field, but not across broad bins of the PPS, suggests
the cluster galaxies are all being preprocessed (e.g., Zabludoff \& Mulchaey 1998, Moran et al. 2007, Kautsch et al. 2008, 
Dressler et al. 2013, Haines et al. 2013, Lopes et al. 2014, Cybulski et al. 2014) in a similar manner during infall, and much of the reduction in star-formation activity has already happened by the time the galaxies enter the broad cluster environment sampled by the OMEGA survey.

We find in Table~\ref{tab:table2} and Section~\ref{sfVagn} that the vast majority
of star-forming galaxies are outside the virialised region (i.e., at large projected radius).
The paucity of star-forming galaxies in the dense virialised region is consistent with studies of other galaxy clusters. 
Many authors have reported a star-formation-density relation at $z\lesssim1$ in which the fraction of star-forming 
galaxies declines from the cluster outskirts (${\rm \sim R_{200}}$) to the cluster core (Kodama et al. 2004, 
Finn et al. 2005, Rines et al. 2005, Koyama et al. 2010, Muzzin et al. 2012, Webb et al. 2013, Wegner et al. 2015).
For example, Kodama et al. (2004) find that the fraction of galaxies detected in H$\alpha$ strongly decreases with local 
density at $z=0.4$. Finn et al. (2005) find no radial trend in star-formation rates within three clusters at $z\sim0.75$. 
They do find, however, that the fraction of star-forming galaxies increases (decreases) with
projected distance (local density) in 2/3 clusters, which is consistent with our study.  

Our results here showing rather small intra-cluster differences in star formation may, at first, seem to be
in conflict with our demonstration in Figures~\ref{env-ksmass50} and \ref{env-bv} of intra-cluster changes in 
galaxy colour. This is not a contradiction because these two results were obtained with different samples. 
The analysis on galaxy colour and other imaging properties involves all 856 galaxies in S2,
whereas the analysis of H$\alpha$-based star formation properties was conducted with the significantly reduced
subset of 351 \textit{emission-line (primarily star-forming)} galaxies in S3 (Figure~\ref{ssfr-def}). In other words, the environmental  changes we have found in the colour distributions refer to the overall galaxy population, while any putative changes in star-formation properties are only relevant to star-forming galaxies. 

\begin{figure*}
\begin{center}
\scalebox{0.85}{\includegraphics[trim={0cm 0cm 0cm 0cm}]{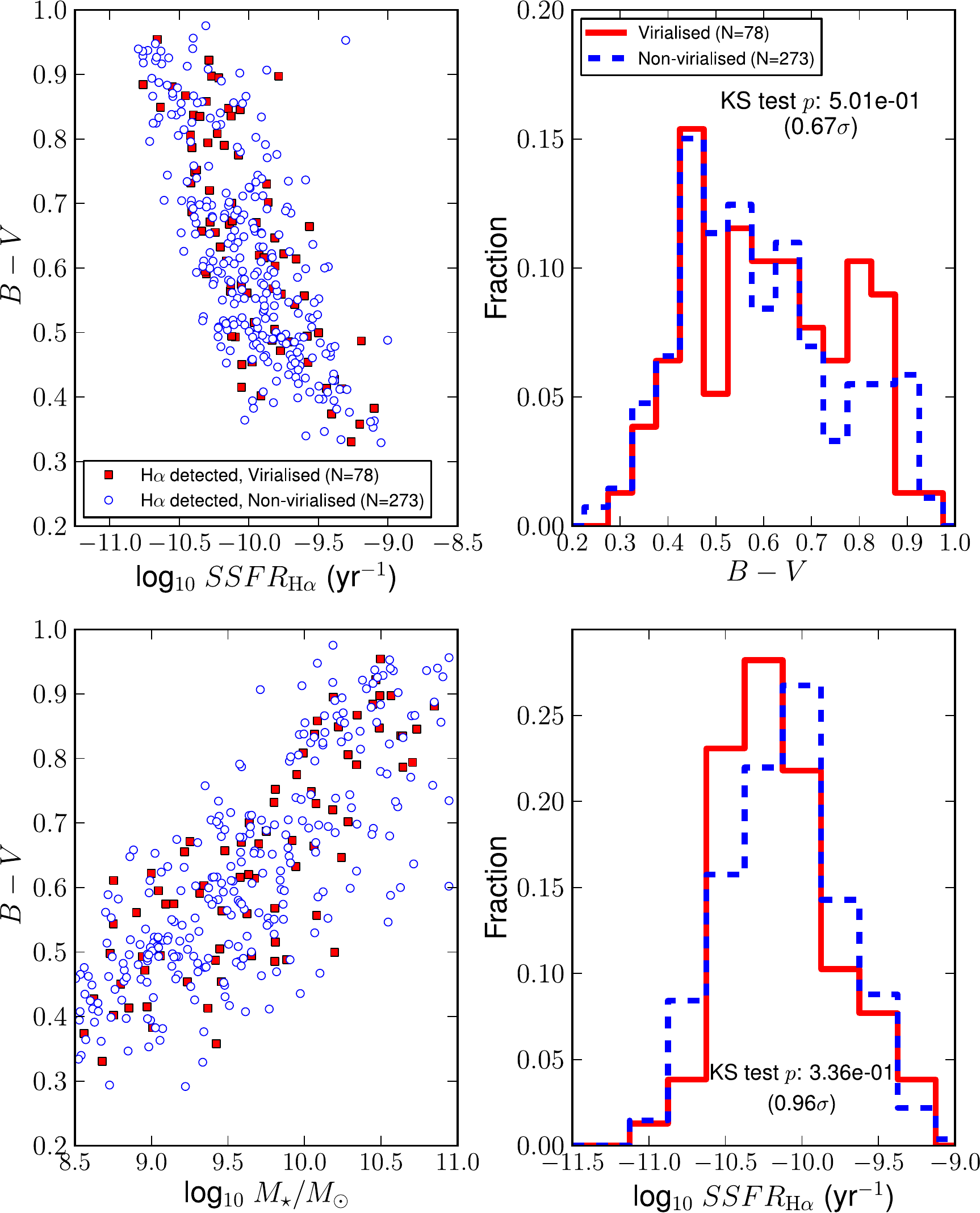}}
\caption{This figure compares $SSFR_{\rm H\alpha}$, colour, and stellar mass for the 351 H$\alpha$-emitting galaxies
in sample S3 (defined in Figure~\ref{ssfr-def}). The points are colour-coded by galaxy position in 
the PPS, as defined by fiducial boundary B1.
\label{ssfr-color-mass}}
\end{center}
\end{figure*}

This is made explicit in Figure~\ref{ssfr-color-mass}, where we compare $SSFR_{\rm H\alpha}$, $B-V$,
and stellar mass for the smaller S3 sample of H$\alpha$-selected galaxies.  The left column shows colour is very similar
at both fixed $SSFR_{\rm H\alpha}$ and mass. The right column shows the histograms of colour and
$SSFR_{\rm H\alpha}$ are not significantly different for the virialised and non-virialised H$\alpha$-emitting galaxies.

\subsection{Properties of Visually Classified Elliptical Galaxies}\label{discuss3}
It is well known that elliptical and early type galaxies are relatively more abundant in dense environments than spiral 
galaxies (e.g., Dressler 1980; Postman \& Geller 1984; Norberg et al. 2002; Goto et al. 2003; Blanton et al. 2005; 
Postman et al. 2005; Wolf et al. 2007; Ball, Loveday \& Brunner 2008). Figure~\ref{env-ksmass50} and 
Table~\ref{tab:table3}, though, show no variation in Hubble type across the PPS within A901/2. Furthermore, at fixed 
Hubble type (E, S0, Spirals, Irr), we have checked for differences in galaxy stellar mass, rest-frame colour, and 
S\'ersic index across the PPS. Visually classified elliptical galaxies are the only population showing marginally 
significant ($\sim2\sigma$) changes in 
colour and stellar mass. No Hubble type shows a substantial change in S\'ersic index across the PPS.

\begin{figure*}
\begin{center}
\scalebox{0.43}{\includegraphics[trim={0.9cm 0 2cm 0}]{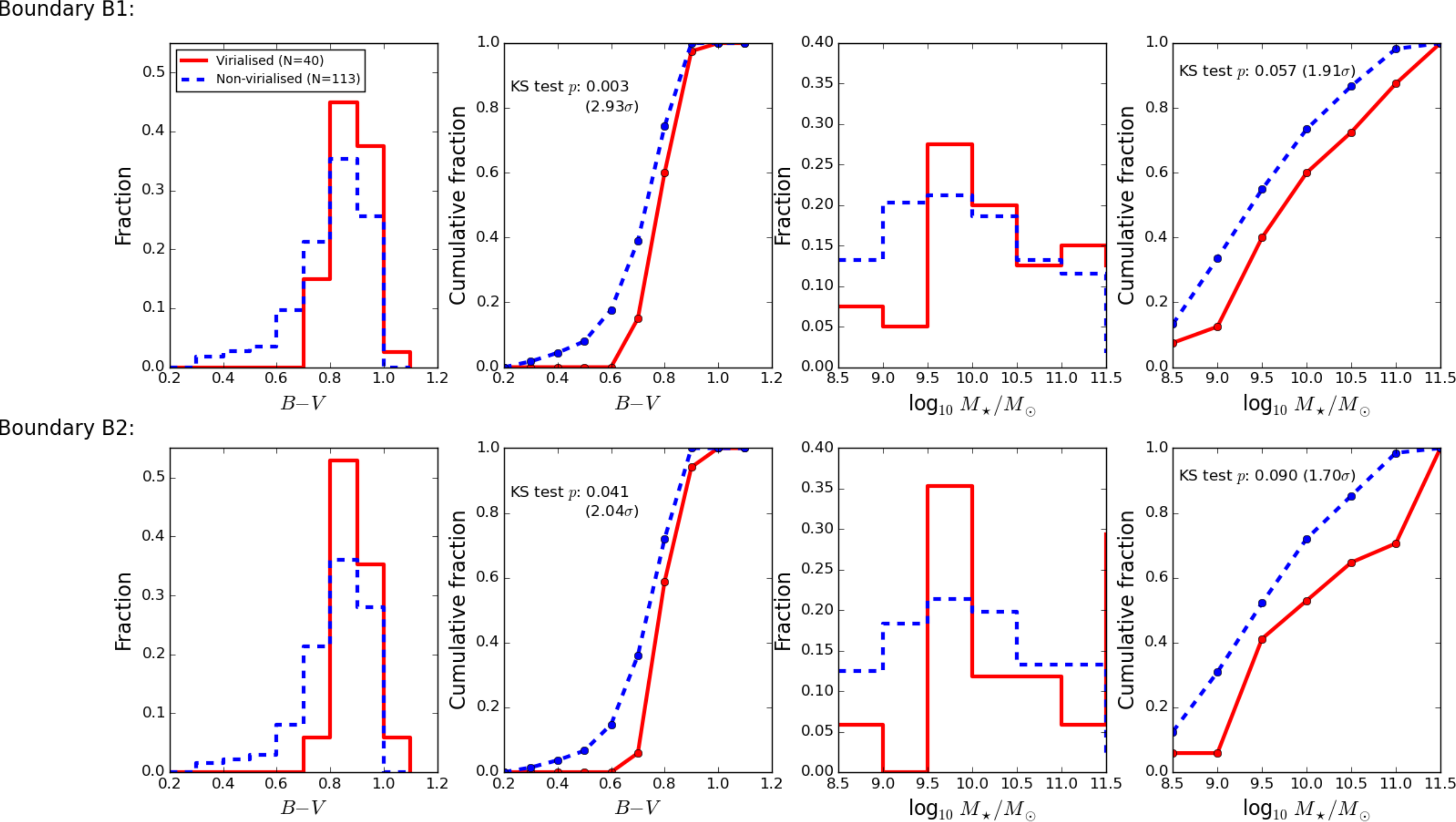}}
\caption{This figure explores how stellar mass and rest-frame colour change for elliptical
galaxies across the PPS. Top row: The distributions of rest-frame $B-V$ and stellar mass are
shown for the two PPS bins (delineated by boundary B1), along with the corresponding cumulative 
distributions. Bottom row: The same, except the PPS bins are defined by boundary B2.
\label{discuss-E}}
\end{center}
\end{figure*}

Figure~\ref{discuss-E} shows the distributions of colour and stellar mass for the elliptical galaxies.  The 
colour distributions of ellipticals are different between the 
PPS bins at the $>2\sigma$ level. This is true for either definition of the virialised region, although 
the difference is more significant for boundary B1 than B2 ($2.9\sigma$ versus $2.0\sigma$). The offset is 
driven by a population of infalling elliptical galaxies with $B-V<0.7$ that is absent in the virialised region. The 
difference in stellar mass due to the PPS, by comparison, is weaker ($1.9\sigma$ for B1 and $1.7\sigma$ for B2), 
indicating the change in colour cannot be explained only by the difference in stellar mass.

A possible caveat with the Hubble types used in this study is that some galaxies called 
``elliptical'' may not be true elliptical galaxies. Studies in the Virgo cluster
show that most true elliptical galaxies are brighter than $M_{\rm V}=-18$ (e.g., Kormendy et al. 2009). 
In our Figure~\ref{env-bv}, however, most of the ellipticals with $B-V<0.7$ are fainter than 
$M_{\rm V}=-18$.  Based on visual inspection, we are confident these objects are not dwarf irregulars. 
Many of them could instead be spheroidal galaxies, defunct
late-types transformed by environment (e.g., Kormendy et al. 2009, Kormendy \& Bender 2012). The
median S\'ersic index ($n=1.77$) of the blue ($B-V\leq 0.7$), visually classified ellipticals is compatible with
this idea.  The redder ($B-V>0.7$) ellipticals, in comparison, have higher median S\'ersic index ($n=3.56$).

\subsection{Differences Between the A901/2 Subclusters}\label{discuss4}
It is interesting to consider how galaxy properties differ across the PPS within the four
A901/2 subclusters (Table~\ref{tab:table1}). Table~\ref{tab:table4} summarises
the effect of the PPS on galaxy properties for each subcluster. Comparing Table~\ref{tab:table1}
and Table~\ref{tab:table4} shows no correlation between subcluster halo properties and the contrast
of galaxy properties across phase-space. It is possible that spatial incompleteness
(Figure~\ref{radec}), especially for A902 and the SW Group, is a limiting factor in the
comparison of individual subclusters.

For the PPS boundary B1, colour and SED type are the only properties consistently varying more strongly than
stellar mass across the PPS in Table~\ref{tab:table4}. Stellar mass changes by $>2\sigma$ only for 
the two more massive subclusters A901a and A901b. These two
subclusters are apparently driving the stronger $3.3\sigma$ change found for stellar mass in
Figure~\ref{env-ksmass50} and Table~\ref{tab:table3}.  The individual subclusters show changes in colour
at the $1.9-3.2\sigma$ level. Except for the SW Group, the subclusters all show environmental changes in
SED type at the $2.2-2.8\sigma$ level. 

A similar contrast in galaxy properties across phase-space is found for the more conservative PPS boundary B2. 
This again highlights the robustness of our conclusions to the particular choice of boundary for the virialised 
region. Colour changes in all subclusters more strongly than stellar mass. This is true for SED type in only 3/4 
subclusters, with A902 being the exception.

\begin{table*}
\footnotesize
  \centering
  \caption{Changes in galaxy properties across PPS and subcluster}
  \label{tab:table4}
  \begin{tabular}{lcccccc}
& & & & \\
\hline
Cluster & Stellar Mass & Colour ($B-V$) & S\'ersic index  & Half-light radius & Hubble type & SED type \\
\hline
\multicolumn{7}{|c|}{\textbf{Boundary B1$^a$}} \\
A901a&3.28$\times 10^{-2}$ (2.13$\sigma$)&1.59$\times 10^{-2}$ (2.41$\sigma$)&1.98$\times 10^{-2}$ (2.33$\sigma$)&1.09$\times 10^{-1}$ (1.60$\sigma$)&1.13$\times 10^{-1}$ (1.58$\sigma$) &2.65$\times 10^{-2}$ (2.22$\sigma$) \\
A901b & 2.01$\times 10^{-2}$ (2.32$\sigma$) & 1.37$\times 10^{-3}$ (3.20$\sigma$)&7.18$\times 10^{-2}$ (1.80$\sigma$)&4.98$\times 10^{-1}$ (0.68$\sigma$)&8.18$\times 10^{-3}$ (2.64$\sigma$)&4.54$\times 10^{-3}$ (2.84$\sigma$)\\
A902 & 4.96$\times 10^{-1}$ (0.68$\sigma$) & 4.16$\times 10^{-3}$ (2.87$\sigma$) & 7.54$\times 10^{-1}$ (0.31$\sigma$) & 9.66$\times 10^{-1}$ (0.04$\sigma$) & 9.58$\times 10^{-1}$ (0.05$\sigma$) & 6.36$\times 10^{-3}$ (2.73$\sigma$) \\
SW Group & 2.08$\times 10^{-1}$ (1.26$\sigma$) & 6.50$\times 10^{-2}$ (1.85$\sigma$) & 2.59$\times 10^{-1}$ (1.13$\sigma$) & 1.30$\times 10^{-1}$ (1.52$\sigma$) & 1.20$\times 10^{-1}$ (1.56$\sigma$) & 4.10$\times 10^{-1}$ (0.82$\sigma$) \\
\hline
\multicolumn{7}{|c|}{\textbf{Boundary B2$^b$}} \\
A901a &7.70$\times 10^{-2}$ (1.77$\sigma$) &3.12$\times 10^{-3}$ (2.96$\sigma$) &2.78$\times 10^{-1}$ (1.09$\sigma$) &4.22$\times 10^{-1}$ (0.80$\sigma$) &5.25$\times 10^{-1}$ (0.64$\sigma$) & 2.99$\times 10^{-5}$ (4.17$\sigma$) \\
A901b &2.40$\times 10^{-1}$ (1.18$\sigma$) &3.00$\times 10^{-2}$ (2.17$\sigma$)&4.89$\times 10^{-2}$ (1.97$\sigma$)&8.79$\times 10^{-1}$ (0.15$\sigma$)&2.92$\times 10^{-1}$ (1.05$\sigma$)&9.07$\times 10^{-3}$ (2.61$\sigma$)\\
A902 & 3.68$\times 10^{-2}$ (2.09$\sigma$) &4.16$\times 10^{-3}$ (2.87$\sigma$) &2.33$\times 10^{-1}$ (1.19$\sigma$) &2.70$\times 10^{-1}$ (1.10$\sigma$) &5.36$\times 10^{-1}$ (0.62$\sigma$) &7.75$\times 10^{-2}$ (1.77$\sigma$) \\
SW Group & 5.89$\times 10^{-1}$ (0.54$\sigma$) & 1.02$\times 10^{-1}$ (1.63$\sigma$) &2.45$\times 10^{-2}$ (2.25$\sigma$) &5.89$\times 10^{-1}$ (0.54$\sigma$) &2.18$\times 10^{-1}$ (1.23$\sigma$) & 8.94$\times 10^{-2}$ (1.70$\sigma$) \\
\hline
\end{tabular}
    \begin{tablenotes}
      \scriptsize 
    \item \textbf{Table notes:}\\
    \item $^{a}$Tests the change in galaxy property inside and outside the virialised region in phase-space defined by boundary B1, $R_{\rm p}$/$R_{200} \leq 1.2$ and 
     $|\Delta V_{\rm los}$/$\sigma_{\rm scl}| \leq 1.5-1.5/1.2 \times R_{\rm p}/R_{200}$.
    \item $^{b}$Tests the change in galaxy property inside and outside a smaller virialised
    region in phase-space defined by boundary B2, $R_{\rm p}$/$R_{200} \leq 0.5$ and 
    $|\Delta V_{\rm los}$/$\sigma_{\rm scl}| \leq 2.0-2.0/0.5 \times R_{\rm p}/R_{200}$.
    \end{tablenotes}
\end{table*}

\section[]{Summary}\label{ssummary}
In this paper, we have conducted a comprehensive, \textit{intra-cluster} analysis of the A901/2 
multi-cluster system at $z\sim0.165$. Aggregating redshifts from traditional spectroscopy, tunable-filter 
imaging, and prism techniques, we have assembled redshifts for a sample of 856 cluster galaxies that reaches in 
stellar mass down to $10^{8.5}$~$M_\odot$. This unique dataset facilitates a more nuanced study of environment 
that goes beyond common field-versus-cluster comparisons. We have looked for variations in cluster galaxy 
properties between the virialised and non-virialised regions of projected phase-space, using two different 
boundaries for the virialised region. Our main results, summarised below, highlight that A901/2 exhibits only 
relatively gentle environmental effects that act mainly on galaxy gas reservoirs.

\begin{enumerate}
\item Stacking the four subclusters (Table~\ref{tab:table1}) in A901/2 together, we find some 
significant changes
across the PPS between the virialised and non-virialised regions for imaging-derived
galaxy properties (Table~\ref{tab:table3}, Figure~\ref{env-ksmass50}, and Figure~\ref{env-chi2}).  
With our fiducial definition of the virialised region (Boundary B1), stellar mass, rest-frame 
$B-V$ colour, and SED type all differ between regions with $>3\sigma$ significance
such that galaxies in the virialised region are more massive, have redder colours, and have
predominantly passive SED types. Additionally, S\'ersic indices are larger, implying dynamically 
hotter stellar structures in the virialised region, at the 2.6$\sigma$ level. Half-light radius 
and Hubble type variations across the PPS are less significant. Similar results are found with 
a smaller, more conservative boundary (B2) for the virialised region.

\item Since stellar mass varies across phase-space in conjunction with other galaxy 
properties, it is important to consider whether 
the apparent variation of these galaxy properties with phase-space is just a mass effect.
Indeed, KS test $p$-value significances for galaxy properties binned in stellar mass are
tens of orders of magnitude stronger than the corresponding test statistics for galaxy properties
binned just in PPS (Table~\ref{tab:table3}, Column 3). 

\item The role of stellar mass is disentangled by conducting the PPS analysis in bins
of stellar mass. At $M_\star > 10^{9.85}$~$M_\odot$, the 50th-percentile in stellar mass, 
the change in rest-frame colour across the PPS is stronger ($3.2\sigma$ versus $0.90\sigma$) than the change in 
stellar mass across the PPS (Figure~\ref{env-ksmass50} and Table~\ref{tab:table3}). This is
due to a population of relatively blue ($B-V<0.7$) galaxies with mostly spiral Hubble types in the higher 
mass bin that exists in
the non-virialised region but not the virialised region. The
prominence of the colour disparity relative to the stellar mass change is even apparent in the
individual subclusters (Table~\ref{tab:table4}). This suggests the changes seen in 
rest-frame colour and SED type are driven to some degree by environmental processes.

\item While Hubble type does not change significantly across the PPS, visually classified elliptical 
galaxies as a group are slightly bluer with $\sim 2\sigma$ significance in the non-virialised region 
(Figure~\ref{discuss-E}). This is due to an infalling population of lower-mass 
($M_\star \leq 10^{9.85}$~$M_\odot$) ellipticals with $B-V<0.7$ that is absent in the 
virialised region. It is important to note that some of the low-luminosity ($M_{\rm V}>-18$), blue ($B-V<0.7$) 
galaxies responsible for this result could be spheroidal galaxies rather than true ellipticals.

\item The proportions of bona-fide star-forming and AGN galaxies change much less in the PPS
than with stellar mass. Still, there is a reduction in the frequency of star-forming galaxies in the virialised region 
(Table~\ref{tab:table2}) that is consistent with the star-formation-density relation
observed in other clusters. 

\item The
{\rm [N\textsc{ii}]}/H$\alpha$ metallicity and $SSFR_{\rm H\alpha}$ deficit differ at $<2\sigma$ 
significance across the PPS. We do, however, see at fixed mass an overall reduction in 
$SSFR_{\rm H\alpha}$ 
relative to the field (Figure~\ref{ssfr-def}), a result found in parallel by Rodr\'iguez del Pino et al. (2017). 
These observations suggest that preprocessing of galaxies during infall plays a prominent
role in quenching star formation.
\end{enumerate}

Our study of galaxy properties across projected phase-space in A901/2 provides compelling evidence for the action
of environmental processes. Since the strongest observable changes in galaxy properties pertain to colour and SED 
type, the environmental processes must be gentle and manifest primarily on the gas reservoirs rather than the
underlying stellar structures. Similar conclusions have been reached for A901/2 previously by (e.g., 
B{\"o}sch et al. 2013; Maltby et al. 2015) and most recently by Rodr\'iguez del Pino et al. (2017) through a 
complementary analysis.

Additional progress in disentangling galaxy environment effects is possible with spatially resolved measures of
star formation and kinematics from integral field units. This will allow a
study in which the spatial distributions and kinematics of young/old stars and
ionised gas can be compared consistently for galaxies spanning a wide range of
environment, stellar mass, star formation, and morphology. 
Studies of this nature are already possible with
data from the Sydney-AAO Multi-Object Integral Field Spectrograph (SAMI, Croom et al. 2012) and Mapping Nearby Galaxies at Apache Point Observatory (MaNGA, Bundy et al. 2015) galaxy surveys.

\section*{Acknowledgements}
Based on observations made with the Gran Telescopio Canarias, installed in the Observatorio del Roque de los Muchachos of the Instituto de Astrof\'isica de Canarias, in the island of La Palma. The GTC reference for this programme is GTC2002-12ESO. Access to GTC was obtained through ESO Large Programme 188.A-2002.
Partly based on observations from the European Southern Observatory Very Large Telescope (ESO-VLT), observing run ID 384.A-0813.
B.R.P acknowledges financial support from the Spanish Ministry of Economy and Competitiveness through grant ESP2015-68964. C.W. was supported by Australian Research Council Laureate Grant FL0992131.

{}

\label{lastpage}

\end{document}